\documentclass[11pt]{article}
\usepackage{jheppub}

\usepackage{epsfig,comment}
\usepackage{graphicx}
\usepackage{url,hyperref}
\usepackage{amsmath,amssymb,bm}
\usepackage{slashed}
\usepackage{float}
\usepackage{orcidlink}
\usepackage{physics,braket}
\usepackage{dsfont}
\usetikzlibrary{quantikz2}
\usepackage{bbold}
\usepackage{tikz}
\usepackage{amsthm}
\usepackage{yfonts}
\usepackage[english]{babel}  

\usepackage{listings}
\usepackage{xcolor}

\usepackage[dvipsnames,table,xcdraw]{xcolor}

\newcommand{\picineq}[1]{\ensuremath{\begin{array}{c} \includegraphics[scale=0.15]{#1} \end{array} } }

\title{Simulating quantum electrodynamics in 2+1 dimensions with qubits and qumodes}

\author[1]{Victor Ale,}
\author[2]{Tommaso Rainaldi,}
\author[3]{Enrique Rico,}
\author[2]{Felix Ringer,}
\author[1]{George Siopsis}

\affiliation[1]{Department of Physics and Astronomy, The University of Tennessee, Knoxville, TN 37996-1200, USA}
\affiliation[2]{Department of Physics and Astronomy, Stony Brook University, New York 11794, USA}
\affiliation[3]{CERN, Theoretical Physics Department, CH-1211 Geneva, Switzerland}

\emailAdd{vale@vols.utk.edu}
\emailAdd{tommaso.rainaldi@stonybrook.edu}
\emailAdd{enrique.rico.ortega@cern.ch}
\emailAdd{felix.ringer@stonybrook.edu}
\emailAdd{siopsis@tennessee.edu}

\abstract{
We develop a hybrid qubit–qumode framework for simulating quantum electrodynamics in 
2+1 dimensions. In this approach, fermionic matter fields are represented by qubits, while 
U(1) gauge fields are encoded in continuous-variable bosonic modes whose canonical quadratures capture the electric and vector-potential components of the theory. To reconcile the non-compact phase space of the qumodes with the compact 
U(1) gauge symmetry, we introduce and compare two complementary constraint-enforcement strategies: (i) a squeezing-based projection that confines qumode states to the unit circle through an effective modification of the inner product, and (ii) a method that dynamically enforces compactness via a penalty Hamiltonian term. We construct the corresponding hybrid Hamiltonian, derive its decomposition into experimentally accessible qubit–qumode gates, and analyze its spectrum in the analytically tractable single-plaquette limit. The hybrid formulation reproduces the correct gauge-invariant dynamics and provides a scalable route toward simulating Abelian lattice gauge theories coupled to fermionic matter on near-term hybrid quantum architectures. Ground-state preparation and convergence are demonstrated using a continuous-variable extension of the Quantum Imaginary Time Evolution (QITE) algorithm, establishing a general framework for hybrid discrete-continuous quantum simulations of lattice gauge theories.}

\begin{document}
\maketitle
\newpage

\section{Introduction}
\label{sec:Introduction}

The real-time dynamics of lattice gauge theories represent one of the most compelling frontiers in quantum simulation, offering access to strongly coupled regimes that remain intractable for classical computation~\cite{banuls2020simulating,PRXQuantum.2.017001,PRXQuantum.4.027001,PRXQuantum.5.037001}. The foundational work of Jordan, Lee, and Preskill demonstrated that scattering processes in scalar field theories can be simulated efficiently with quantum computers, providing a concrete pathway for exploring quantum field dynamics using controllable quantum systems \cite{Jordan:2012xnu}. Building on the Hamiltonian formulation introduced by Kogut and Susskind \cite{Kogut:1974ag}, a variety of quantum algorithms have been proposed for studying Abelian and non-Abelian gauge theories, confinement phenomena, and particle production. Extending these methods to real-time dynamics of quantum electrodynamics (QED) and quantum chromodynamics (QCD) remains a key goal for both high-energy, nuclear, and condensed-matter physics.

Simulating such theories on quantum hardware requires representing both fermionic and gauge-boson degrees of freedom—objects of fundamentally different character. Fermionic fields are naturally discrete and can be mapped efficiently to qubits via transformations such as Jordan–Wigner or Bravyi–Kitaev~\cite{Jordan:1928wi,Bravyi:2000vfj}. Gauge fields, by contrast, are continuous variables living on group manifolds, e.g., U(1), SU(2), typically represented in qubit-based simulations by truncating their Hilbert spaces to finite-dimensional subspaces \cite{Chandrasekharan:1996ih,Zohar:2015hwa,Martinez2016,PhysRevA.98.032331,Kokail2019,Alexandru:2019nsa,Haase:2020kaj,Bauer:2021gek,Ciavarella:2021nmj,Davoudi:2022xmb,Grieninger:2023ehb,Halimeh:2023lid}. While systematic, such truncations often introduce significant resource overhead and discretization errors that grow near the continuum limit. Continuous-variable (CV) quantum systems, whose observables are described by canonical quadratures of bosonic modes \cite{Lloyd:1998jk}, offer a natural alternative for encoding gauge fields without discretization \cite{Marshall:2015mna}. 

A broad spectrum of approaches to quantum simulation of lattice gauge theories has emerged over the past decade. Early digital implementations using trapped ions and superconducting qubits have demonstrated real-time dynamics and string breaking in 
1+1-dimensional gauge models \cite{Martinez2016,Kokail2019}, establishing the viability of Hamiltonian-based digital quantum simulation. Analog and hybrid schemes have been explored in cold-atom systems \cite{PhysRevA.105.023322} and in trapped-ion and circuit-QED architectures capable of realizing gauge-invariant couplings \cite{Muschik_2017}. Continuous-variable (CV) quantum information has recently gained attention as a natural framework for representing bosonic fields without truncation: studies of the O(3) model~\cite{Jha:2023ecu}, scalar field theory~\cite{Briceno:2023xcm,Abel:2024kuv,Abel:2025pxa,Abel:2025zxb,Gupta:2025xti}, SU(2) lattice gauge theory~\cite{Ale2025}, quantum electrodynamics and related field theories~\cite{Crane:2024tlj,miranda2025renormalized,k9p6-c649}, $Z_2$-Higgs~\cite{saner2025real,varona2024towards,domanti2025dynamical,bǎzǎvan2024synthetic,Schuckert:2025iko} and Yukawa \cite{Than:2025gso}. These developments motivate the present work, which integrates discrete qubits for fermionic matter with CV qumodes for gauge fields into a unified hybrid architecture, bridging discrete and continuous paradigms.

Recent advances in experimental architectures now enable hybrid quantum platforms that coherently couple discrete and continuous degrees of freedom, including trapped ions, superconducting circuits, and cavity-QED systems \cite{Stavenger:2022wzz,Liu:2024mbr,Araz:2024dcy,6prx-zmdz}. In trapped-ion devices, qubits are encoded in the internal electronic states of ions, while motional modes act as bosonic qumodes \cite{RevModPhys.75.281,wineland1998experimentalissuescoherentquantumstate}; in superconducting circuits, nonlinear Josephson junctions serve as qubits coupled to harmonic resonators that realize qumodes \cite{Blais:2020wjs}. Such hybrid systems offer access to universal sets of gates combining qubit and qumode operations, including Gaussian gates, non-Gaussian phase gates, and conditional displacements \cite{Liu:2024mbr}. These developments open a promising avenue for hybrid discrete-continuous quantum simulation of field theories, where the discrete and continuous sectors can be encoded natively in the hardware.

In this work, we introduce a hybrid qubit–qumode approach to simulating quantum electrodynamics in 
2+1 dimensions, a paradigmatic Abelian lattice gauge theory that captures key features of gauge-field dynamics in higher dimensions while remaining computationally tractable. In our formulation, fermionic matter fields are represented by qubits, and the U(1) gauge degrees of freedom are encoded in pairs of qumodes. This construction enables a continuous-variable representation of the gauge fields that preserves gauge invariance without explicit truncation. To reconcile the non-compact nature of qumodes with the compact U(1) gauge symmetry, we introduce two complementary constraint-enforcement schemes:
(i) a squeezing-based projection that restricts qumode states to the unit circle through an effective modification of the inner product, and
(ii) a method that dynamically enforces compactness by adding corresponding penalty terms to the Hamiltonian.

We demonstrate both approaches in the single-plaquette model, where analytical and numerical results confirm convergence to the correct U(1) spectrum. The hybrid Hamiltonian is further decomposed into elementary qubit–qumode gates compatible with experimental realizations, enabling direct implementation on near-term hybrid platforms. We also employ a continuous-variable generalization of the Quantum Imaginary Time Evolution (QITE) algorithm for ground-state preparation \cite{Motta2020,PhysRevA.105.012412}, establishing a scalable route toward hybrid simulation of interacting gauge–fermion systems.

The framework developed here extends naturally to larger lattices and to non-Abelian gauge theories, where the combination of discrete and continuous resources may provide an efficient and physically transparent encoding. Beyond applications in high-energy and nuclear physics, similar techniques can be adapted to model fermion–boson interactions in quantum chemistry, condensed-matter, and polaritonic systems \cite{Kang:2023xfb,Vu:2025aub}.

The hybrid qubit–qumode representation developed here is complementary to previous digital or continuous-variable simulations, as it preserves the full infinite-dimensional Hilbert space of the gauge field without resorting to truncation and embeds the compact U(1) structure directly at the operator level. This allows gauge invariance to be maintained exactly while retaining the experimental accessibility of hybrid architectures, providing a distinct alternative to qubit-only encodings of lattice gauge theories.

The remainder of this paper is organized as follows. Section \ref{sec:2} introduces the notation and hybrid gate set. Section \ref{sec:3} formulates the lattice Hamiltonian for U(1) gauge theory in 
2+1 dimensions, and Section \ref{sec:4} describes its quantum implementation with qubits and qumodes. Section \ref{sec:single} presents numerical results for a single plaquette. We conclude in Section \ref{sec:7} with an outlook on scalability and applications of hybrid quantum simulations.

\section{Notation and qubit-qumode gates}
\label{sec:2}

Hybrid quantum simulation combines discrete-variable (DV) qubits with continuous-variable (CV) bosonic modes (``qumodes'' or harmonic oscillators) to form a unified computational basis. Throughout this work, we denote by $\hat{q}$ and $\hat{p}$ the canonical quadrature operators of a qumode satisfying 
\begin{equation}
    [\hat{q}, \hat{p}] = i\ ,
\end{equation}
with units such that $\hbar =1$. Each qumode thus realizes an infinite-dimensional Hilbert space $\mathcal{H}_\text{CV}$, spanned by the eigenstates $\ket{q}$ or $\ket{p}$ of the corresponding quadrature operators.

Qubits are represented by two-level systems with computational basis states $\ket{0}$ and $\ket{1}$.
We denote the Pauli operators by $\sigma \in \{ X,Y,Z \}$, and the ladder combinations
\begin{equation}\label{eq:Xpm}
    X^\pm = \frac{1}{2} (X \pm iY) \ .
\end{equation}
Multi-qubit operators are expressed as tensor products of Pauli strings, $P = \sigma_{i_1}\otimes \sigma_{i_2}\otimes \cdots$, which form a complete operator basis for the discrete sector  $\mathcal{H}_\text{DV}$. 

The combined system of $n_q$ qubits and $n_c$ qumodes resides in
\begin{equation}
    \mathcal{H}_\text{hybrid} = \mathcal{H}_\text{DV} \otimes \mathcal{H}_\text{CV} \ .
\end{equation}
Operators acting on the hybrid space may couple qubits and qumodes.
A general hybrid operator takes the form
\begin{equation}
    \hat{O} = P \otimes f(\hat{q}, \hat{p}) \ , 
\end{equation}
where $P$ is a tensor product of Pauli matrices acting on qubits and $f(\hat{q}, \hat{p})$ is a polynomial or analytic function of the quadratures.
The elementary commutation relations between these operators govern the construction of gate sets for hybrid quantum simulation.

Following Lloyd and Braunstein \cite{Lloyd:1998jk}, universal CV computation is generated by Gaussian gates (displacements, rotations, squeezing) together with a single non-Gaussian operation such as the cubic-phase gate,
\begin{equation}\label{eq:cpg}
    {\rm V}(\gamma) = e^{i\frac{\gamma}{3} \hat{q}^3}\ ,
\end{equation}
as well as the two-qumode beam splitter gate
\begin{eqnarray}
    {\rm BS}(z)= e^{z \hat a^\dagger \hat b- z^* \hat a \hat b^\dagger} \,,
\end{eqnarray}
where $\hat a,\hat b$ refer to the bosonic lowering operators of two different qumodes. When combined with universal qubit gates $\{ R_x, R_y, R_z, \text{CNOT} \}$, hybrid universality can be achieved by introducing at least one entangling qubit–qumode operation.
A convenient choice, experimentally realizable in trapped-ion and superconducting-circuit platforms, is the conditional displacement gate
\begin{equation}
    {\rm CD} (\alpha)  = e^{(\alpha \hat a^\dagger - \alpha^* \hat a) Z},
\end{equation}
where $\hat a = \frac{1}{\sqrt{2}} (\hat{q} + i \hat{p})$ is the bosonic annihilation operator and the Pauli matrix $Z$ acts on a control qubit.

Together, the set
\begin{equation}\label{eq:universalgates}
    \mathcal{G}_\text{hybrid} = \left\{ {\rm R}_{x,y,z} , \text{CNOT}, {\rm D}(\xi), {\rm S}(z) , {\rm V}(\gamma), \text{BS}(z), \rm{CD}(\alpha ) \right\}\,,
\end{equation}
is universal for hybrid computations~\cite{Liu:2024mbr}. Note that this is not a minimal universal gate set since, for example, the cubic phase gate can be constructed from conditional displacements and qubit gates. In addition, different choices of universal gate sets are possible. For example, the conditional displacement gate $\rm{CD}(\alpha)$ in Eq.~(\ref{eq:universalgates}) can be replaced with the red sideband/Jaynes-Cummings gate listed in table~\ref{table:1} to achieve a universal hybrid gate set. See in particular Ref.~\cite{Liu:2024mbr} for more details. All elements of the Hamiltonians studied in this work can be expressed as combinations of these gates or their infinitesimal generators. In our simulation of 
2+1-dimensional quantum electrodynamics, fermionic matter fields are encoded in qubits via a Jordan–Wigner transformation, while gauge fields residing on lattice links are represented by qumodes. Our simulation protocol employs several additional gates. While these additional gates can be decomposed into the elementary operations listed in Eq.~(\ref{eq:universalgates}), we list all relevant gates used in this work in Table~\ref{table:1}.

\begin{table}
    \centering
    \begin{tabular}{c|c|c|c}
        Gate type & Operation & Short & Operator \\ \hline\hline
        Qubit & Pauli operators & & $\sigma^i$ \\
        & Rotation & ${\rm R}_i(\theta)$ & $e^{i\theta\sigma^i/2}$ \\
        & Controlled NOT & CNOT & $e^{i\frac{\pi}{4}(\mathbb{I}_1-Z_{1})(\mathbb{I}_2-X_{2})}$ \\ \hline  
      Qumode & Rotation & ${\rm R}(\theta)$ & $e^{i\theta \hat a^\dagger \hat a}$ \\
      & Fourier & ${\rm F}$ & $e^{i\frac{\pi}{2} \hat a^\dagger \hat a}$ \\
       & Displacement & ${\rm D}(z)$ & $e^{z \hat a^\dagger-z^* \hat a}$\\
       & Single-mode squeezing & ${\rm S}(z)$ & $e^{(z^* \hat a\hat a-z \hat a^{\dagger}\hat a^\dagger)/2}$\\
       & Beam splitter & ${\text{BS}}(z)$ & $e^{z \hat a^\dagger \hat b- z^* \hat a \hat b^\dagger}$  \\
       & Kerr & ${\rm K}(z)$ & $e^{iz (\hat a^\dagger \hat a)^2}$ \\
       & Cross-Kerr & ${\rm CK}(z)$ & $e^{iz \hat a^\dagger \hat a\, \hat b^\dagger \hat b}$ \\
       & Quadratic phase  & ${\rm P}(\theta)$ & $e^{i\frac{\theta}{2} \hat{q}^2}$ \\
       & Cubic phase & ${\rm V}(\theta)$ & $e^{i\frac{\theta}{3} \hat q^3}$ \\
       \hline
        Hybrid & Red sideband/Jaynes Cummings  & ${\rm RSB}(z)$&$  e^{i z \hat a X^+ + i z^* \hat a^\dagger X^-}$ \\
         & Blue sideband/Anti-Jaynes Cummings & ${\rm BSB}(z)$&$e^{i z \hat a^\dagger X^+ + i z^* \hat a X^-}$ \\
         &Controlled rotation & ${\rm CR}(\theta)$&$e^{i\theta Z \hat a^\dagger\hat a}$\\ 
         &Controlled displacement & ${\rm CD}(z)$&$e^{Z(z \hat a^\dagger-z^* \hat a)}$ \\ 
         &Controlled squeezing & ${\rm CS}(z)$&$e^{Z(z^* \hat a\hat a-z \hat a^{\dagger}\hat a^\dagger)/2}$\\ 
         &Controlled beam splitter & ${\rm CBS}(z)$&$e^{Z (z \hat a^\dagger \hat b- z^* \hat a \hat b^\dagger)}$ \\ \hline
    \end{tabular}
    \caption{Hybrid qubit/qumode gates for the simulation of QED in 2+1 dimensions. The gates are parametrized in terms of a complex variable $z=\theta e^{i\phi}$ with $\theta\ge 0$ and $\phi\in[0,2\pi)$.}
    \label{table:1}
\end{table}

The gates above are accessible in several hybrid architectures:
\begin{itemize}
    \item \underline{Trapped ions and neutral atoms}: Qubits can be realized in terms of different electronic states or specifically hyperfine splittings of states. Qumodes can be realized in terms of vibrational modes of the ions or atoms. For example, for trapped ions, collective vibrational modes of the ion chain are particularly useful to realize qumodes due to their resilience to environmental noise. Conditional displacements can be realized by spin-dependent forces. See Refs.~\cite{porras2004, serafini2009, debnath2018,Chen_2021, Katz:2022gra, bouchoule1999, shaw2024, bohnmann2024}.
\item
\underline{Superconducting circuits}: Qubits are encoded in the nonlinear degrees of freedom of Josephson junctions, such as transmons or fluxonium qubits, while qumodes are realized in terms of electromagnetic modes of microwave resonators or cavities. The nonlinearity of the Josephson element enables tunable qubit–qumode couplings, including cross-Kerr interactions and conditional displacements that serve as building blocks for hybrid gates. See Refs.~\cite{Blais:2004hcj,Heeres:2015cnj,Blais:2020wjs,Eickbusch:2021uod,Liu:2024mbr}.
\item
\underline{Cavity QED and optomechanical systems:} These platforms realize light-matter interactions through radiation-pressure or dispersive couplings. In cavity-QED systems, a two-level system interacts with a cavity photon mode, forming a qubit/qumode system. Optomechanical systems couple a cavity field to a mechanical oscillator, providing a purely continuous-variable system. See Refs.~\cite{Hatridge:2013xmi,Higginbotham:2018mzw,PhysRevA.75.063401}.
\end{itemize}
These physical primitives motivate the gate-level decompositions developed later, but are not yet tied to a specific field-theoretic mapping. Depending on the hardware platform, qubits or qumodes may exhibit longer or shorter coherence times. The connectivity, native interaction strengths, and the set of experimentally accessible gates can also vary significantly, which can impact the construction of hybrid quantum algorithms.

The definitions above provide the mathematical and operational background for hybrid computations.
In Section \ref{sec:3}, we turn to the independent physics of the U(1) gauge theory in 2+1 dimensions, deriving its Hamiltonian formulation, symmetries, and spectrum without reference to any encoding.
The hybrid representation of these operators and their simulation on quantum hardware will then be introduced in Section \ref{sec:4}.

\section{$U(1)$ lattice gauge theory with fermions in 2+1 dimensions}\label{sec:3}

Before constructing quantum simulation protocols, we first summarize the physics of lattice quantum electrodynamics (QED) in 2+1 dimensions. In this section, we describe the Hamiltonian formulation of an Abelian U(1) gauge theory coupled to fermionic matter~\cite{Kogut:1974ag}, including its electric and magnetic components, local gauge symmetry, and Gauss’s law constraints. We derive the reduction of degrees of freedom imposed by gauge invariance, analyze the exactly solvable single-plaquette case, and obtain the normal-mode spectrum for extended lattices.
These results establish the physical quantities (energy scales, excitation gaps, and constraint structures) that will be encoded and simulated using the hybrid qubit–qumode framework introduced in section \ref{sec:4}.

\subsection{Hamiltonian formulation}

The Hamiltonian approach to lattice gauge theory provides a natural framework for describing the real-time dynamics of gauge fields coupled to fermionic matter. For an Abelian U(1) gauge theory discretized on a two-dimensional spatial lattice with spacing 
$a$, the Hamiltonian separates into electric and magnetic components \cite{Kogut:1974ag}
\begin{equation}
        H_E = \frac{g^2}{2}\sum_{\bm{n}} E^2_{\bm{n},i}\ ,\ \
    H_B  = -\frac{1}{2g^2a^2} \sum_{\bm{n}}\left(P_{\bm{n}}+P^{\dagger}_{\bm{n}}-2\right) \, ,
    \label{eq:pure_U1_Hamiltonian}
\end{equation}
where ${\bm n}\equiv (n_x,n_y)$ labels lattice sites, $i\in \{ x,y \}$ denotes link directions, $g$ is the gauge coupling with dimension 
$[g]=1/2$, and $a$ is the lattice spacing with dimension $[a]=-1$. We will work with units where $a=1$.

The operator $E_{\bm{n},i}$ represents the lattice
electric field on the link emanating from site $\bm{n}$ in direction $i$, and its conjugate link variable
\begin{equation}
    U_{\bm{n},i} = e^{i\chi_{\bm{n},i}} \in U(1),
\end{equation}
encodes the vector potential.
They satisfy the canonical commutation relation
\begin{equation}\label{eq:canstr}
\left[E_{\bm{n},i},U_{\bm{n}',j}\right] = \delta_{\bm{n}\bm{n}'}\delta_{ij} U_{\bm{n},j} \ ,
\end{equation}
which mirrors the continuum relation between the electric field and vector potential. The magnetic part of the Hamiltonian is written in terms of oriented products of link variables around each plaquette,
\begin{equation}\label{eq:Pn}
    P_{\bm{n}}\equiv U_{\bm{n},x}U_{\bm{n}+ \bm{e}_x, y}U^{\dagger}_{\bm{n}+\bm{e}_y,x} U^{\dagger}_{\bm{n},y} \ ,
\end{equation}
so that $H_B$ reproduces the lattice curl of the vector potential. In the continuum limit, it approaches the familiar expression $H_B \to \frac{1}{2} \int d^2 x B^2$, with $B = \partial_x A_y - \partial_y A_x$.

Charged fermionic matter fields residing on lattice sites can be incorporated through the Kogut–Susskind (staggered) formulation \cite{Kogut:1974ag}.
Their addition entails two more contributions to the Hamiltonian:
\begin{equation}
     H_M  = m_0\sum_{\bm{n}}(-)^{n_x + n_y}\Psi^{\dagger}_{\bm{n}}\Psi_{\bm{n}} \ ,\ \
    H_K  = \frac{1}{2} \sum_{\bm{n}}\sum_{i \in \{ x,y\} } \left[\Psi^{\dagger}_{\bm{n}}U^{\dagger}_{\bm{n},\bm{e}_{i}}\Psi_{\bm{n}+\bm{e}_{i}} + \text{h.c.}\right] \ ,
\end{equation}
where $m_0$ is the bare mass of the fermionic field $\Psi_{\bm{n}}$. 

The full Hamiltonian is therefore
\begin{equation}\label{eq:Hfull}
    H = H_E + H_B + H_M + H_K \ .
\end{equation}
This model, quantum electrodynamics in 
2+1 dimensions, retains the essential interplay between electric and magnetic fields, and includes massive charged fermions that can form bound states or particle–antiparticle pairs.
In the continuum limit, $m_0 , g\to 0$, and the number of plaquettes diverges. The Hamiltonian in the continuum is labeled by a single dimensionless parameter, $m_0/g^2$.

The full Hamiltonian \eqref{eq:Hfull} is gauge invariant, commuting with the generators $G_{\bm{n}}$ of local gauge transformations,
\begin{equation}\label{eq:Gn}
    G_{\bm{n}}\equiv \sum_{i\in \{ x,y\} }\left(E_{\bm{n},i} - E_{\bm{n}+\bm{e}_i,i} \right)-Q_{\bm{n}},
\end{equation}
where $Q_{\bm{n}}$ is the charge operator at lattice site $\bm{n}$. In the staggered fermion formulation, it is given by
\begin{equation}
    Q_{\textbf{n}} = \Psi^{\dagger}_{\bm{n}}\Psi_{\bm{n}} - \frac{1-(-)^{n_x+n_y}}{2} \mathbb1.
\end{equation}
The Hilbert space of physical states is restricted by imposing the Gauss law constraints, 
\begin{equation}
    G_{\bm{n}}\ket{\text{phys}} = 0\,.
\end{equation}
The total charge vanishes for systems with open boundaries and neutral backgrounds,
\begin{equation}\label{eq:Qtot}
    Q_{\text{total}} = \sum_{\bm{n}} Q_{\bm{n}} = 0 \ .
\end{equation}
It is possible to add non-dynamical background charges $\mathcal{Q}_{\bm{n}}$ by modifying Gauss's law to $G_{\bm{n}}\ket{\text{phys}} = \mathcal{Q}_{\bm{n}}\ket{\text{phys}}$, thus altering the physical content of the system in the continuum. Here, we will set all background charges to zero. It is straightforward to add them to our discussion. 

The constraints drastically reduce the number of independent degrees of freedom.
For a lattice of $(2N+1)\times (2N+1)$ plaquettes, the total number of links is 
$4(N+1)(2N+1)$, while there are 
$4(N+1)^2 -1$ independent Gauss-law conditions.
Hence, the number of physical gauge degrees of freedom equals $(2N+1)^2$, matching the number of plaquettes -- one dynamical variable per plaquette corresponding to the magnetic flux through that cell.

We emphasize that all results in this Section are derived independently of any quantum encoding; the expressions for the Hamiltonian, Gauss-law constraints, and plaquette structure provide the physical foundation upon which the hybrid representation of Section \ref{sec:4} will act.

\subsection{Single plaquette}
\begin{figure}[t]
    \centering
        \includegraphics[width=0.7\linewidth]{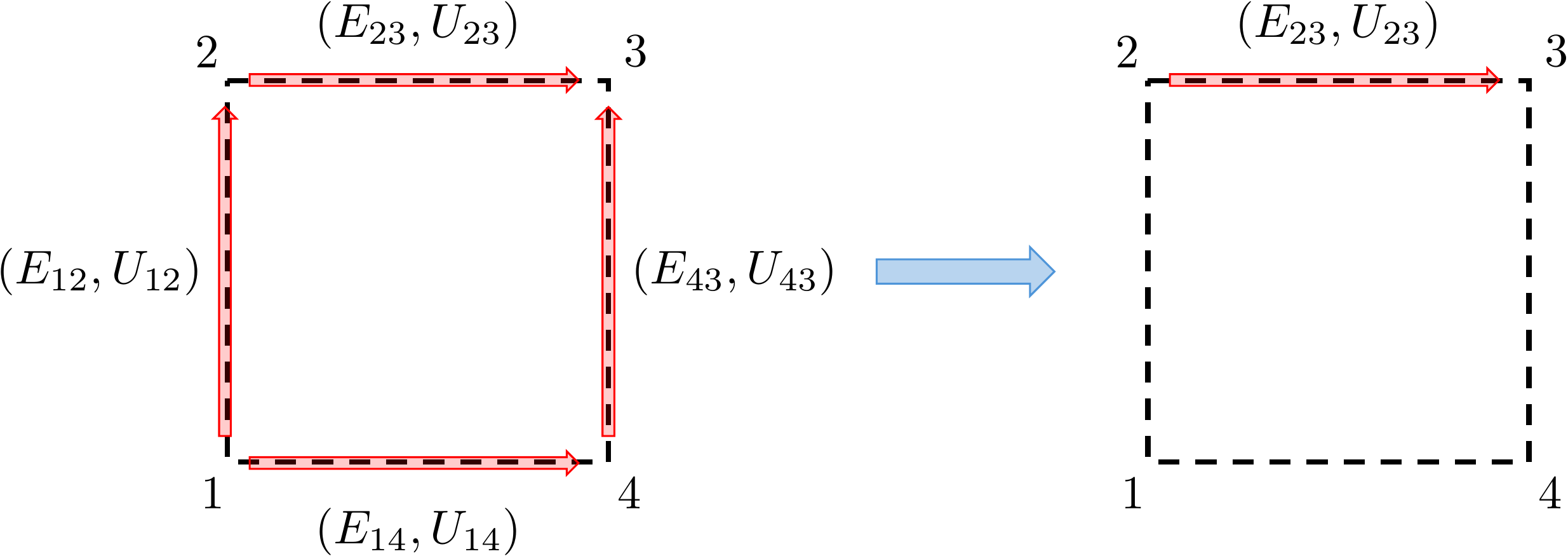}
    \caption{One plaquette system for the pure $U(1)$ gauge theory before (left) and after (right) the selection of the dynamical degrees of freedom.}
    \label{fig:One_plaquette_U1}
\end{figure}

The simplest system we can study is a single plaquette consisting of four links and four vertices in a chargeless background. For simplicity, we label the four vertices as $(0,0)\equiv 1, (0,1)\equiv 2, (1,1)\equiv 3$ and $(1,0)\equiv 4$. The electric field and link variables are relabeled according to Fig.~\ref{fig:One_plaquette_U1}.
Gauss's law at the four vertices imposes the constraints
\begin{equation}
    \begin{split}
        E_{12}+E_{14}&=Q_1,\\
        E_{23}-E_{12} & = Q_2,\\
        -E_{23}-E_{43} & = Q_3,\\
        E_{43}- E_{14} & =Q_4,.
    \end{split}
\end{equation}
Notice that they are compatible only if the total charge vanishes, see Eq.\ \eqref{eq:Qtot},
\begin{equation}\label{eq:qt}
   Q_{\mathrm{total}} = Q_1+Q_2+Q_3+Q_4 = 0\,.
\end{equation}
Thus, there is only one independent gauge degree of freedom. Selecting the pair $(E_{23},U_{23})$ to be the dynamical gauge degree of freedom of the single plaquette system, we arrive at the electric and magnetic parts of the Hamiltonian,
\begin{equation}
    \begin{split}
        H_E & = \frac{g^2}{2}\left[E^2 + \left(E-Q_2\right)^2 + \left(E+Q_3\right)^2 + \left(E - Q_1 - Q_2\right)^2\right],\\
    H_B & = -\frac{1}{2g^2}\left(U + U^{\dagger}-2\right) = \frac{1}{g^2} (1 - \cos\chi)\,.\\
    \end{split}
\end{equation}
Here we simplified the notation by denoting $E\equiv E_{23}$ and $U\equiv U_{23} = e^{i\chi}$. The two additional contributions from fermions are
\begin{equation}
    \begin{split}
        H_M & = m_0\sum_{i=1}^4(-1)^{1+i}\Psi^{\dagger}_i\Psi_i,\\
    H_K & = \frac{1}{2} \left(\Psi^{\dagger}_1\Psi_2 + \Psi^{\dagger}_1\Psi_4 + \Psi_2U^{\dagger}\Psi^{\dagger}_3 + \Psi_4\Psi^{\dagger}_3\right) + \text{h.c.}\,.
    \end{split}
\end{equation}
Fermions live in a 16-dimensional Hilbert space. However, the requirement of a vanishing total charge restricts the physical space to 6 dimensions.

In the static charge limit (large $m_0$), the fermionic Hamiltonian is dominated by $H_M$.  The fermionic Hilbert space is spanned by 16 basis vectors which can be obtained by repeatedly acting with modes $\Psi_i, \Psi_i^\dagger$ ($i=1,2,3,4$) on the vacuum state $\ket{\Omega_f}$ which is annihilated by $\Psi_{1,3} , \Psi_{2,4}^\dagger$.
There are 6 states with $Q_{\mathrm{total}} =0$,
\begin{equation}\label{eq:wff}
        \begin{array}{cccccc}
        \picineq{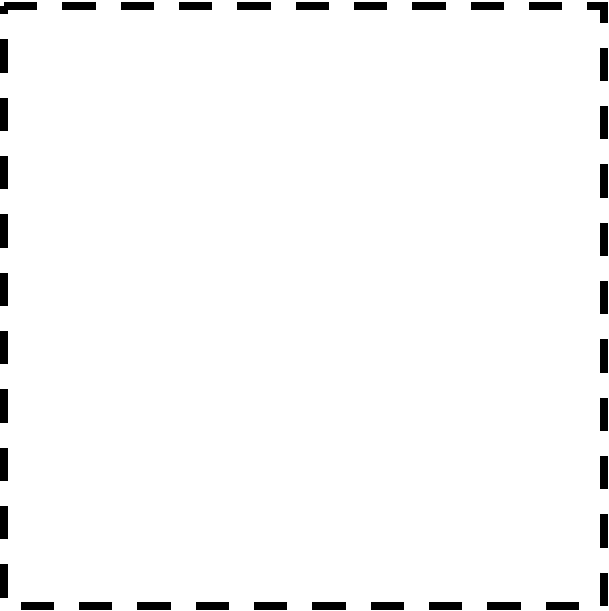} &  & \picineq{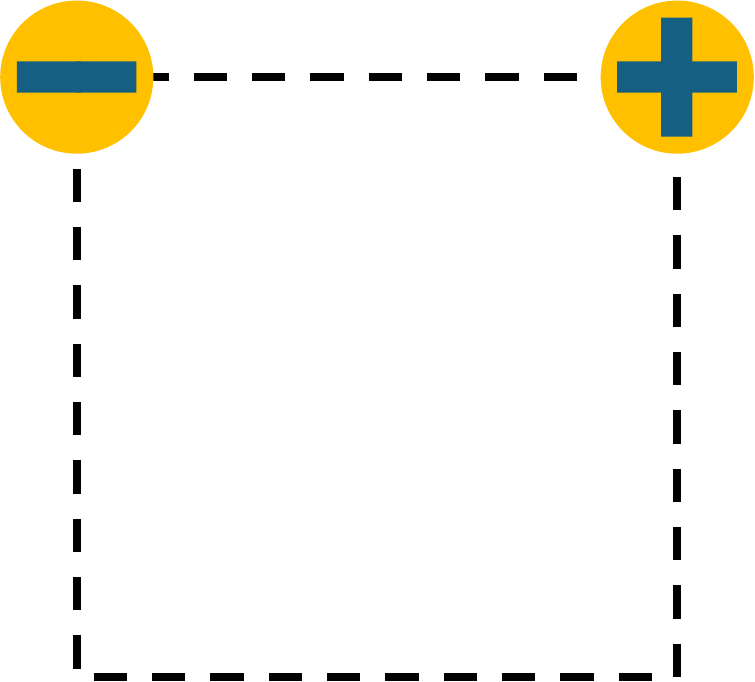} &  & \picineq{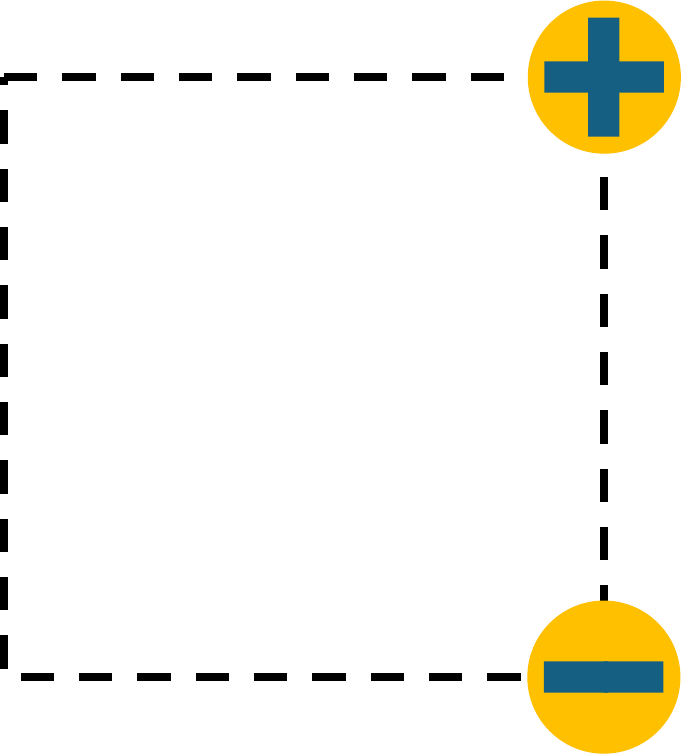} & \\  \ket{v_1} \equiv \ket{\Omega_f}   & , & 
         \ket{v_2} \equiv \Psi_2 \Psi_3^\dagger \ket{\Omega_f} & , & \ket{v_3} \equiv \Psi_3^\dagger \Psi_4 \ket{\Omega_f}  & , \\
        =\ket{vvvv}  & & =\ket{ve^-e^+v} & &=\ket{vve^+e^-} & \\  \picineq{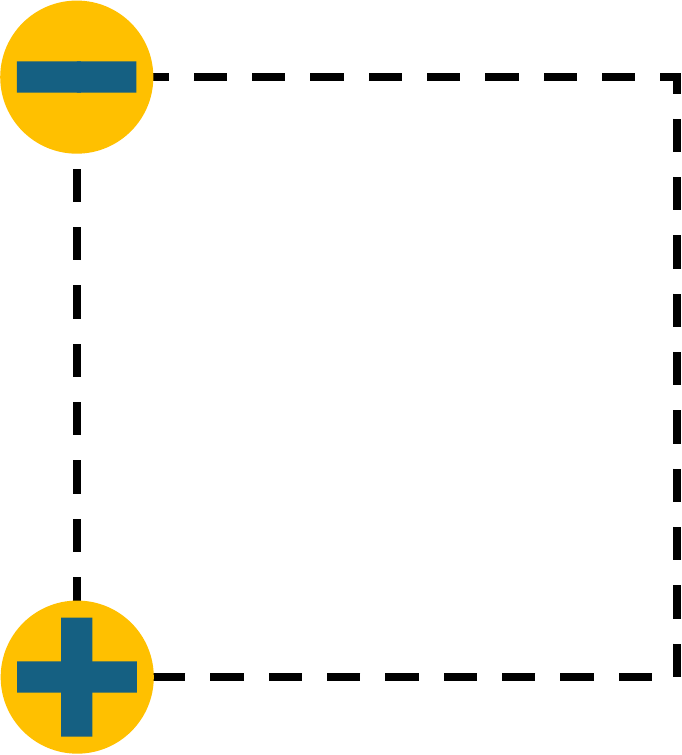} & & \picineq{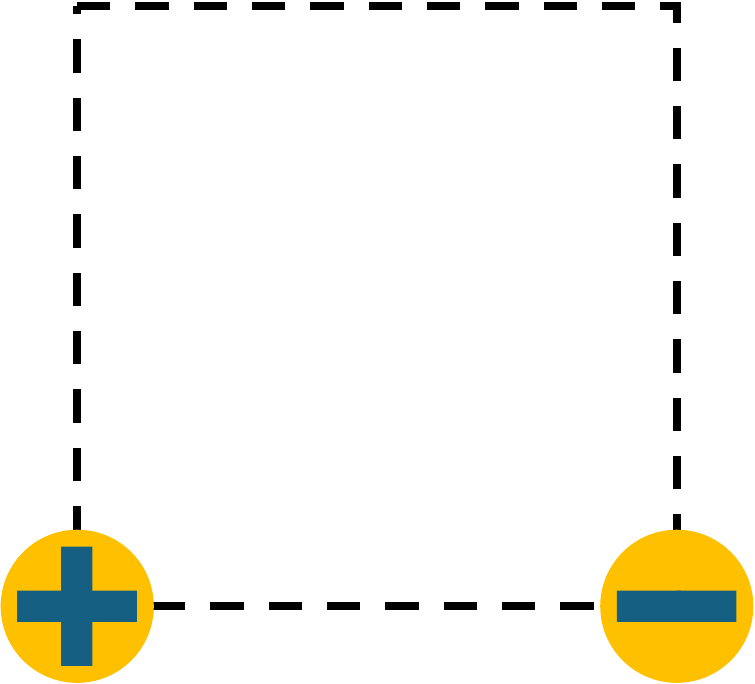} & & \picineq{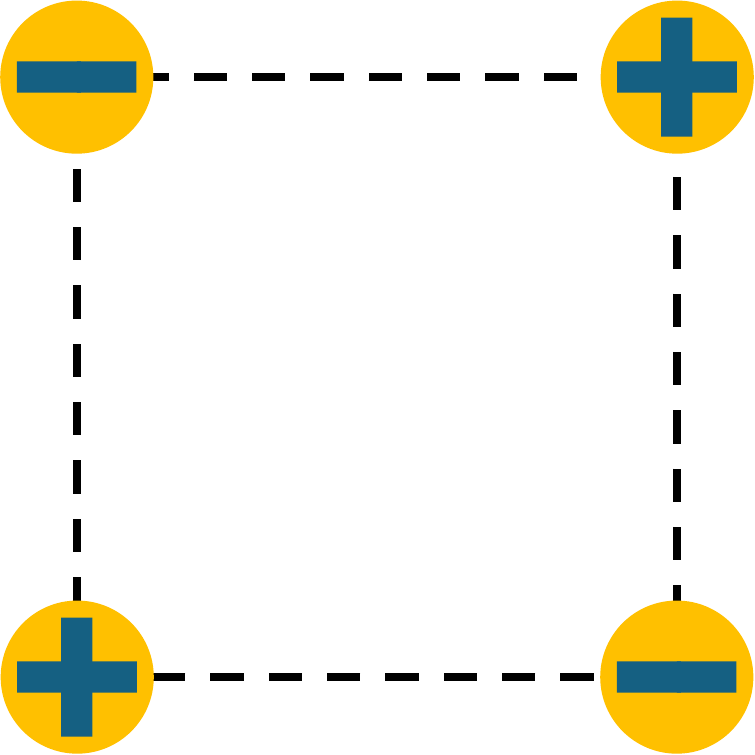} & \\ \ket{v_4} \equiv \Psi_1^\dagger \Psi_2 \ket{\Omega_f} & , &  \ket{v_5} \equiv \Psi_1^\dagger \Psi_4 \ket{\Omega_f}  &  ,& \ket{v_6} \equiv \Psi_1^\dagger \Psi_2 \Psi_3^\dagger \Psi_4\ket{\Omega_f}  & . \\ =\ket{e^+e^-vv}  & & =\ket{e^+vve^-}  & & =\ket{e^+e^-e^+e^-} & \end{array}
\end{equation}
Their energies, the eigenvalues of $H_M$, are $(-2m_0, 0, 2m_0)$, with a four-fold degenerate zero eigenvalue corresponding to a particle-antiparticle pair.
Their charges are
\begin{equation}
Q_{1,3}\ket{e^+} = \ket{e^+},\quad Q_{1,2,3,4}\ket{v}=0,\ 
     Q_{2,4}\ket{e^-}=-\ket{e^-}\,,
\end{equation}
in accord with the staggered fermion prescription.

The full Hamiltonian 
\begin{equation}
    H = H_E + H_B + H_M + H_K\,,
\end{equation}
contains a term which is linear in $E$, which is of the form
\begin{equation}
    -g^2 \mathcal{A} E \ , \ \ \mathcal{A} = Q_1 + 2Q_2 -Q_3\,.
\end{equation}
It can be eliminated by performing the unitary transformation
\begin{equation}
    \mathcal{U} = e^{\frac{i}{4} \mathcal{A} \chi}\,,
\end{equation}
which commutes with the charges and does not affect $H_M$. After some algebra, we obtain
\begin{equation}
    \widetilde{H} \equiv \mathcal{U}^\dagger H \mathcal{U} =   \widetilde{H}_g + \widetilde{H}_f + \widetilde{H}_K\,,
\end{equation}
where
\begin{eqnarray}
    \widetilde{H}_g &=& 2g^2 E^2 + \frac{1}{g^2} (1-\cos\chi) \nonumber\\
   \widetilde{H}_f &=& H_M + \frac{g^2}{8} \left( 3(Q_1+Q_3) - 4 Q_2Q_4 -  Q_1Q_3 \right)  \nonumber\\
    \widetilde{H}_K &=& \frac{1}{2} e^{\frac{i}{4} \chi} \left(  \Psi^{\dagger}_1\Psi_2 +  \Psi^{\dagger}_4\Psi_1 +  \Psi_3\Psi^{\dagger}_2 +  \Psi_4\Psi^{\dagger}_3\right)  + \text{h.c.}
\end{eqnarray}
The eigenstates of the gauge part of the Hamiltonian, $\widetilde{H}_g$,
\begin{equation}
    \widetilde{H}_g \ket{\Phi_{n,g}} = E_{n,g} \ket{\Phi_{n,g}} \ ,
\end{equation}
can be written analytically in terms of Mathieu functions ($n=0,1,2,\dots$)
\begin{equation}
    E_{n,g} = \begin{cases}
        \frac{g^2}{2}a_n\left(-\frac{1}{g^4}\right) +\frac{1}{g^2},\quad &n\,\text{even},\\
        \frac{g^2}{2}b_{n+1}\left(-\frac{1}{g^4}\right) +\frac{1}{g^2},\quad &n\,\text{odd},\\
    \end{cases}\quad\longleftrightarrow\quad \Phi_{n,g} (\chi) \propto \begin{cases}
        \text{ce}_n\left(\frac{\chi}{2},-\frac{1}{g^4}\right),\quad &n\,\text{even},\\
        \text{se}_{n+1}\left(\frac{\chi}{2},-\frac{1}{g^4}\right),\quad &n\,\text{odd}.
    \end{cases}
    \label{eq:1plaquette_U1_spectrum}
\end{equation}
The eigenstates of the static fermionic part, $\widetilde{H}_f$, are given by Eq.\ \eqref{eq:wff},
\begin{equation}
    \widetilde{H}_f \ket{v_i} = E_{i,f} \ket{v_i}\,,
\end{equation}
where
\begin{equation}
    E_{1,f} = -2m_0 \ , \ E_{2,f} =   E_{3,f} =  E_{4,f} =  E_{5,f} = \frac{3g^2}{8} \ , \ E_{6,f} = 2m_0 + \frac{g^2}{8}\,.
\end{equation}
In the static limit, we can ignore $\widetilde{H}_K$. We deduce the zeroth-order energy levels
\begin{equation}
    E_{n,i}^{(0)} = E_{n,g} + E_{i,f} \quad\longleftrightarrow\quad \ket{\Psi_{n,i}^{(0)}} = \mathcal{U} \ket{\Phi_{n,g}} \otimes \ket{v_i}\,.
\end{equation}
The ground state is 
\begin{equation}
\ket{\Psi_{0,1}^{(0)}} = \int_{-\pi}^\pi d\chi \, \text{ce}_0\left(\frac{\chi}{2},- \frac{1}{g^4} \right) \ket{\chi}  \otimes \ket{vvvv}\,,
\end{equation}
and its energy is $E_{0,1}^{(0)} = \frac{g^2}{2} a_0 (1/g^4) + 1/g^2 -2m_0$. The first excited state is 
\begin{equation}
\ket{\Psi_{1,1}^{(0)}} = \int_{-\pi}^\pi d\chi \, \text{se}_2\left(\frac{\chi}{2},- \frac{1}{g^4} \right) \ket{\chi}  \otimes \ket{vvvv}\,,
\end{equation}
with energy $E_{1,1}^{(0)} = \frac{g^2}{2} b_2 ( \frac{1}{g^4} ) + \frac{1}{g^2} -2m_0$. We obtain an energy gap of $\Delta E = E_{1,1}^{(0)} - E_{0,1}^{(0)} \sim \mathcal{O} (1)$.
\begin{figure}[t]
    \centering
    \includegraphics[width=\linewidth]{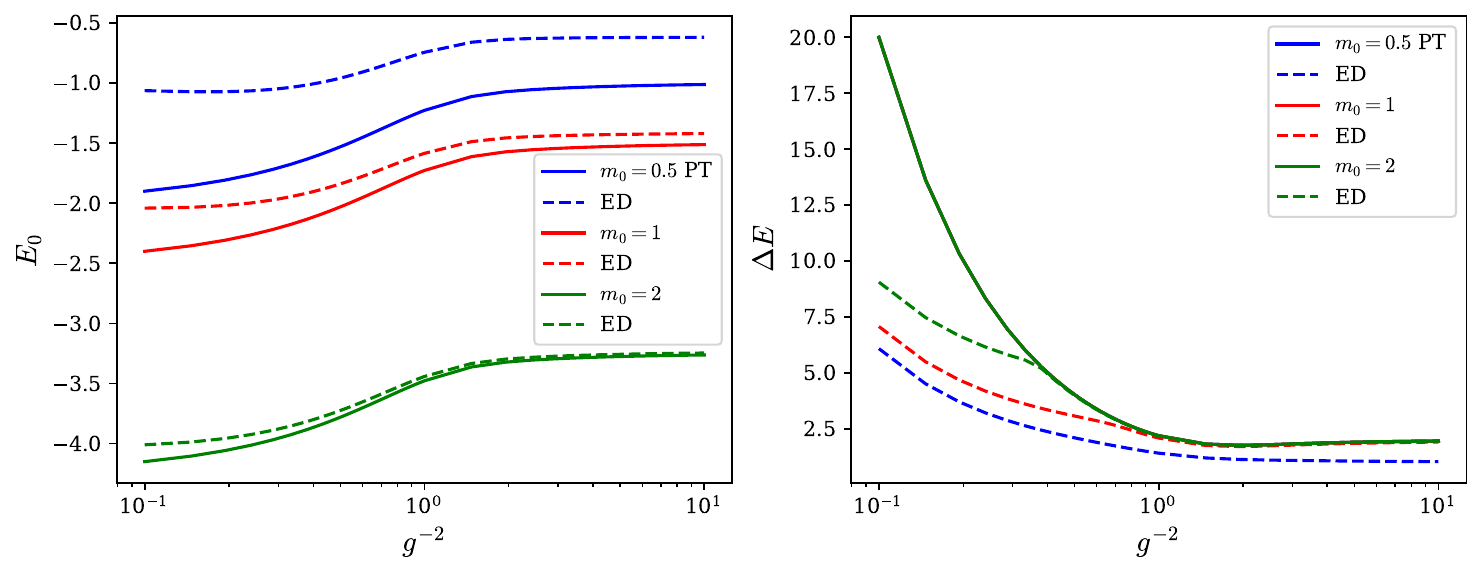}
    \caption{Comparison of the perturbative analysis (PT) in Eq.~\eqref{eq:grndE} with exact diagonalization (ED). We show the ground state energy (left) and the energy gap (right) for different values of the mass $m_0$.}
    \label{fig:PTED}
\end{figure}
Away from the static limit, we can treat ${H}_K$ as a perturbation which amounts to an expansion in $m_0^{-1}$.
For the ground state, first-order perturbation produces no shift in energy. At the second order, we obtain the energy level
\begin{equation}
    E_{0} = E_{0,1}^{(0)} - \sum_{(n,i)\ne (0, 1)} \frac{\vert \bra{\Psi_{0,1}^{(0)}} H_K \ket{\Psi_{n,i}^{(0)}} \vert^2}{E_{n,i}^{(0)} - E_{0,1}^{(0)}} + \mathcal{O} (m_0^{-2})\,.
\end{equation}
After some algebra, we obtain
\begin{equation}\label{eq:grndE}
    E_0 = -2m_0 + \frac{g^2}{2} a_0 \left( \frac{1}{g^4} \right) + \frac{1}{g^2} - \frac{1}{2m_0}  + \mathcal{O} (m_0^{-2})\,.
\end{equation}
A comparison of the perturbative results and exact diagonalization  (ED) is shown in Fig.~\ref{fig:PTED}.  As expected, the two results agree in the large mass region.

\subsection{Square lattice}

Next, we generalize the above results to a square lattice with $(2N+1)\times (2N+1)$ plaquettes. We consider a lattice with open boundary conditions. This system has $4(N+1)^2$ lattice sites and $4(2N+1)(N+1)$ links. Each link accommodates a gauge degree of freedom. These degrees of freedom are constrained by Gauss's Law at each lattice site. However, we have $4(N+1)^2 -1$ independent constraints due to the vanishing total charge of the system. Therefore, the number of independent gauge degrees of freedom is $4(2N+1)(N+1)-4(N+1)^2 +1 = (2N+1)^2$. Let $\bm{n} = (n_x,n_y)$ label a lattice site with $n_x, n_y=0,1,\dots,2N+1$. We will fix the gauge so that all link variables are set to the identity $\mathbb{1}$ except the horizontal link variables $U_{\bm{n},x} = e^{i\chi_{\bm{n},x}} \in U(1)$ with $n_x, n_y=0,1,\dots,2N$ and $\chi_{\bm{n},x} \in [-\pi , \pi)$, which are chosen as our $(2N+1)^2$ independent gauge degrees of freedom. Using Gauss's law \eqref{eq:Gn}, we can express the electric field in terms of the independent fields $E_{\bm{n}, x}$, which are conjugate to the link variables,
\begin{equation}
    [ E_{\bm{n}, x}, U_{\bm{n}', x}] = \delta_{\bm{nn}'} U_{\bm{n}, x}\ , \ \ [ E_{\bm{n}, x}, \chi_{\bm{n}', x}] = -i\delta_{\bm{nn}'}\,.
\end{equation}
We deduce the electric and magnetic parts of the Hamiltonian,
\begin{eqnarray}\label{eq:HEB}
    H_E &=& g^2\sum_{\bm{n},\bm{n}'} \left( \mathcal{H}_{\bm{n}\bm{n}'}^{(2)} E_{\bm{n}, x} E_{\bm{n}', x} + \mathcal{H}_{\bm{n}\bm{n}'}^{(1)} Q_{\bm{n}} E_{\bm{n}',x} + \mathcal{H}_{\bm{n}\bm{n}'}^{(0)} Q_{\bm{n}} Q_{\bm{n}'} \right)\,, \nonumber\\
    H_B &=& \frac{1}{g^2} \sum_{\bm{n}} \left( 1 - \cos (\chi_{\bm{n}, x} - \chi_{\bm{n} +e_y, x}) \right)\,.
\end{eqnarray}
For the electric part, $H_E$, we sum over all lattice sites. Since we are working with $E_{\bm{n}, x}$ for $n_x, n_y < 2N+1$, we have $\mathcal{H}_{\bm{n}\bm{n}'}^{(2)} = 0$, if one of $n_x, n_y, n_x', n_y'$ is $2N+1$. Similarly, $\mathcal{H}_{\bm{n}\bm{n}'}^{(1)} = 0$, if one of $n_x=2N+1$, or $n_y =2N+1$. For the magnetic part, $H_B$, the sum runs over lattice sites with $n_x, n_y < 2N+1$.

After gauge fixing, the fermionic contributions are
\begin{eqnarray}\label{eq:Hf0}
    H_M &=& m_0 \sum_{\bm{n}} (-)^{n_x+n_y} \Psi_{\bm{n}}^\dagger \Psi_{\bm{n}} \ ,\nonumber\\
    H_K &=&  \frac{1}{2} \sum_{\bm{n}}  \Psi_{\bm{n}}^\dagger U^\dagger_{\bm{n}, x} \Psi_{\bm{n}+ e_x} + \frac{1}{2}\sum_{\bm{n}}   \Psi_{\bm{n}}^\dagger \Psi_{\bm{n}+ e_y}  + \mathrm{h.c.} \,.
\end{eqnarray}
We consider the weak-coupling and heavy-fermion limit, where we can obtain explicit analytic expressions. The weak coupling limit is obtained by rescaling $E_{\bm{n}, x} \to  E_{\bm{n}, x}/g$, $\chi_{\bm{n}, x} \to g \chi_{\bm{n}, x}$. We also expand in $1/m_0$. At leading order, the Hamiltonian is
\begin{equation}
    H =  H_g^{(0)} + H_M + {\cal O}(1)\,,
\end{equation}
where the gauge field contribution is
\begin{equation}
    H_g^{(0)} = \sum_{\bm{n},\bm{n}'} \mathcal{H}_{\bm{n}\bm{n}'}^{(2)} E_{\bm{n}, x} E_{\bm{n}', x} + \frac{1}{2} \sum_{\bm{n}} \left( \chi_{\bm{n}, x} - \chi_{\bm{n} +e_y, x} \right)^2 \ .
\end{equation}
Thus, the bosonic gauge part consists of $(2N+1)^2$ coupled harmonic oscillators. By diagonalizing the matrix $\mathcal{H}_{\bm{n}\bm{n}'}^{(2)}$, we obtain the frequencies of the normal modes,
\begin{equation}
    \omega^g_{\bm{k}} = 2\sin^2 \frac{\pi k_x }{4(N+1)} + 2\sin^2 \left( \frac{(k_x+1)\pi}{2} + \frac{\pi k_y }{4(N+1)} \right) \ , \ \ k_x,k_y = 1,\dots, 2N+1 .
\end{equation}
The gap in the bosonic sector is given by the lowest normal frequency, $4\sin^2 \frac{\pi}{4(N+1)}$, obtained by setting $k_x=k_y=1$,
\begin{equation}
    \Delta E = E_1 - E_0 = 4\sin^2 \frac{\pi}{4(N+1)} + {\cal O}(1) \ .
\end{equation}
As expected, the gap vanishes in the continuum limit ($N\to\infty$).

For the fermionic part, we obtain modes with frequencies $\omega^f = \pm m_0$. 
We associate positive frequency modes with particles and negative frequency modes with antiparticles. 

This completes the field-theoretic description of the 
U(1) gauge theory coupled to fermionic matter in 2+1 dimensions. The next section translates these dynamics into the language of hybrid quantum simulation, specifying how electric, magnetic, and matter operators are encoded and implemented using qubits and qumodes.

\section{Quantum computation with qubits and qumodes}
\label{sec:4}

Having established the field-theoretic structure of 
U(1) gauge theory coupled to fermions in 2+1 dimensions, we now formulate a representation suitable for quantum simulation on hybrid architectures that combine discrete (qubit) and continuous (qumode) degrees of freedom. We express the full lattice Hamiltonian in terms of experimentally accessible qubit–qumode gates. This formulation will serve as the foundation for the numerical results presented in the subsequent section.

\subsection{Hybrid representation of lattice operators}

Each fermionic site 
$\bm{n} = (n_x,n_y)$ is represented by a qubit whose computational basis encodes the local occupation number. In accord with the staggered fermion prescription, the single site state $\ket{0}_{\bm{n}}$ represents a particle of charge $+1$ for odd $n_x+n_y$ and $\ket{1}_{\bm{n}}$ is the vacuum state for these sites whereas for even $n_x+n_y$, the vacuum is represented by $\ket{0}_{\bm{n}}$ and an anti-particle with charge $-1$ by $\ket{1}_{\bm{n}}$.
In this representation, the charge operators become projection operators acting on single qubits,
\begin{equation}
        Q_{\bm{n}} \mapsto  \frac{Z_{\bm{n}} + (-)^{n_x+n_y} \mathbb{1}}{2}\ .
\end{equation}
The fermionic creation and annihilation operators are mapped to qubit operators as
\begin{equation}
    \Psi_{\bm{n}}^\dagger \mapsto X_{\bm{n}}^+ \ , \ \ \Psi_{\bm{n}} \mapsto X_{\bm{n}}^- \ ,
\end{equation}
where $X^\pm$ is defined in \eqref{eq:Xpm}, with appropriate Jordan–Wigner strings ensuring fermionic anticommutation relations across sites.

Gauge fields residing on links $(\bm{n}, i)$ with $i\in \{ x,y \}$ are represented by pairs of qumodes with quadratures $\bm{q}_{\bm{n}, i} = (q_{\bm{n}, i}^0,q_{\bm{n}, i}^1)$ and $\bm{p}_{\bm{n}, i} = (p_{\bm{n}, i}^0,p_{\bm{n}, i}^1)$, obeying the commutation rules 
\begin{equation}\label{eq:43}
    [q_{\bm{n}, i}^\mu, p_{\bm{n}', j}^\nu] = i\delta_{\bm{nn}'} \delta_{ij} \delta^{\mu\nu} \ .
\end{equation}
To connect the field-theoretic formulation outlined in Section \ref{sec:3} with a representation suitable for quantum simulation, we now promote the classical lattice variables $(E_{\bm{n},i}, U_{\bm{n},i})$
to operators acting on hybrid qubit-qumode registers. The mapping below defines the correspondence between the canonical variables of the lattice theory and their quantum counterparts realized on the qumode Hilbert space:
\begin{equation}\label{eq:44}
    E_{\bm{n},i} \mapsto J_{\bm{n},i} \equiv q_{\bm{n},i}^0 p_{\bm{n},i}^1 - q_{\bm{n},i}^1 p_{\bm{n},i}^0 \ , \ \   U_{\bm{n},i} \mapsto q_{\bm{n},i}^0 + iq_{\bm{n},i}^1 \ .
\end{equation}
This correspondence identifies the lattice electric field $E_{\bm{n},i}$ with the angular momentum operator $J_{\bm{n},i}$ in the two-dimensional space of the pair of bosonic modes and the link operator $U_{\bm{n},i}$ with a complexified coordinate. This identification ensures that local gauge transformations act on the hybrid operators in the same way as in the classical theory. Indeed, the unitary implementing gauge transformations is
\begin{equation}
    \mathcal{U}_{\rm{gauge}} (\bm{\theta}) = \prod_{\bm{n}} e^{i\theta_{\bm{n}} \mathcal{G}_{\bm{n}}} \ , \ \ \mathcal{G}_{\bm{n}} = \sum_{i\in \{ x,y \}} (J_{\bm{n}, i} - J_{\bm{n} + e_i, i} ) - Q_{\bm{n}} \,,
\end{equation}
\emph{cf.}\ Eq.\ \eqref{eq:Gn}. Using the canonical commutation relations of the qumode quadratures \eqref{eq:43}, we obtain
\begin{align}
    \mathcal{U}_{\rm{gauge}} (\bm{\theta}) \left( q_{\bm{n},i}^0 + iq_{\bm{n},i}^1 \right) \mathcal{U}_{\rm{gauge}}^\dagger (\bm{\theta}) = &\,e^{i(\theta_{\bm{n}} - \theta_{\bm{n} + e_i})} \left( q_{\bm{n},i}^0 + iq_{\bm{n},i}^1 \right) \, , \nonumber\\
    \,  \mathcal{U}_{\rm{gauge}} (\bm{\theta}) J_{\bm{n},i} \,\mathcal{U}_{\rm{gauge}}^\dagger (\bm{\theta}) = &\, J_{\bm{n},i} \, ,
\end{align}
reproducing the transformation properties of the classical link variables. Thus the dictionary in Eq.\ \eqref{eq:44} not only maps the lattice Hamiltonian to hybrid quantum operators but also guarantees gauge invariance at the operator level, ensuring that all subsequent algorithmic steps act within the gauge-invariant subspace.

By employing two qumodes for each link, we have enlarged the phase space. However, we have not broken gauge invariance because the electric field acts as angular momentum that leaves the unit circle $\bm{q}_{\bm{n},i}^2 =1$ invariant, $[ J_{\bm{n},i} , \bm{q}_{\bm{n},i}^2 ] =0$. Consequently, $\bm{q}_{\bm{n},i}^2$ is conserved as it commutes with the Hamiltonian. To restore compactness of the link variables $U_{\bm{n},i}$, we ought to impose the constraints
\begin{equation}\label{eq:uc}
    \bm{q}_{\bm{n},i}^2 = (q_{\bm{n},i}^0)^2 + (q_{\bm{n},i}^1)^2 = 1\ .
\end{equation}
We introduce two complementary methods to restore compactness within this CV representation.

\subsubsection*{Compactification Method A}

We create compact continuous variables through a squeezed-state projection that modifies the inner product in the qumode Hilbert space. Let $\Lambda$ be the effective squeezing parameter (to be defined precisely below). Wavefunctions are suppressed outside the unit circles \eqref{eq:uc}, yielding expectation values that converge to the desired compact-group results as $\sim \Lambda^{-2}$. In the limit $\Lambda \to\infty$, each qumode effectively wraps onto the unit circle, realizing $U_{\bm{n},i}$ as a true phase operator.

\begin{figure}[ht!]
    \centering
\begin{quantikz}
    \lstick{$\mu$} & \gate[2]{\text{BS}^{\dagger}} & \gate{\text{V} (\sqrt{2}s)} & \gate[2]{\text{BS}} & &\\
    \lstick{$a$} & & \gate{\text{V} (-\sqrt{2}s)} & & \gate{\text{V} (s)} &
\end{quantikz}
    \caption{Circuit implementing the non-Gaussian unitary $\mathcal{D}_\mu (s)$ given in Eq.\ \eqref{eq:ds}.}
    \label{circ:1}
\end{figure}
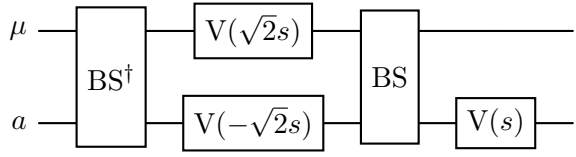

The gauge part of any state that we consider must be of the form
\begin{equation}\label{eq:82}
    \ket{\psi^{{(Q)}}} \propto \int d^2 q\, \delta (\bm{q}^2 -1) \psi(\bm{q}) \ket{\bm{q}},
\end{equation}
where the delta function enforces the unitarity constraint in Eq.~\eqref{eq:uc}. We denote with the superscript $(Q)$ any state or operator used in the quantum simulation. To implement this state, we apply a non-Gaussian unitary $\mathcal{D}(s)$ that entangles each of the system qumodes $q_\mu$ ($\mu=0,1$) with an ancilla prepared in the quadrature eigenstate $\ket{p_a =s}_a$, where $s$ is a free parameter. This unitary is defined as
\begin{equation}\label{eq:ds}
    \mathcal{D}_\mu (s) = e^{-is q_{\mu}^2 q_a}\,.
\end{equation}
It can be implemented with beam splitters and cubic phase gates \eqref{eq:cpg} as
\begin{equation}\mathcal{D}_\mu (s) = \text{BS}_{\mu a}^{\dagger}\ \text{V}_a(-\sqrt{2}s) \cdot \text{V}_{\mu}(\sqrt{2}s)\ \text{BS}_{\mu a}\cdot \text{V}_a(s)\end{equation}
The circuit implementing this is shown in Fig.~\ref{circ:1}.
Then the circuit that creates $\ket{\psi^{(Q)}}$ from input $\ket{\psi} = \int d^2 q\, \psi (\bm{q}) \ket{\bm{q}}$ is given in Fig.~\ref{circ:2}.
\begin{figure}[ht!]
    \centering
\begin{quantikz}
    \lstick{$\ket{p=0}_a$} & & \gate{\text{D} (is)} & & \gate{\mathcal{D}(s)} & \gate{\mathcal{D}(s)} & \meterD{p=0}\\
    \lstick[2]{$\ket{\psi}$} &   & & &\ctrl{-1} & &  \\
    &  & &  & & \ctrl{-2} & 
\end{quantikz}
    \caption{The quantum circuit that creates the state  $\ket{\psi^{(Q)} }$, see Eq.\ \eqref{eq:82}, from a given input two-qumode state $\ket{\Psi}$, where D is a displacement gate (see Table \ref{table:1}) and $\mathcal{D}$ is defined in Eq.\ \eqref{eq:ds}.}
    \label{circ:2}
\end{figure}
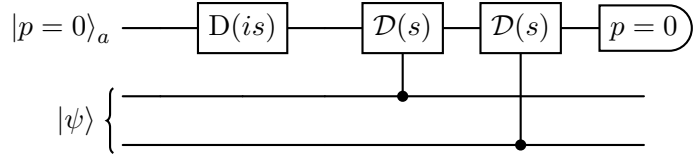

In a realistic implementation, the initial state of the ancilla qumode in the quantum circuit of Fig.\ \ref{circ:2}  is approximated by a squeezed state $S(r) \ket{0}_a$, where $S(r) = e^{\frac{r}{2} (a^2 - {a^\dagger}^2)}$ is the squeezing operator. The modified quantum circuit yields the state
\begin{equation}\label{eq:82m}
    \ket{{\psi}_{\mathrm{approx}}^{{(Q)}} (\Lambda)} \propto \int d^2 q\, e^{- \frac{\Lambda^2}{g^2} (\bm{q}^2 -1)^2} \psi(\bm{q}) \ket{\bm{q}}, \ \Lambda = \frac{g}{2} s e^r
\end{equation}
which reduces to the desired state $\ket{\psi^{(Q)}}$ in Eq.\ \eqref{eq:82} in the limit of infinite squeezing, $r\to\infty$.
The unitarity constraint $\delta (\bm{q}^2 -1)$ has been traded for a sharply peaked Gaussian. 

\subsubsection*{Compactification Method B}

An alternative way of enforcing compactness is by adding a penalty term to the Hamiltonian,
\begin{equation}\label{eq:LM}
    H_\mu = \frac{\mu}{2} \sum_{\bm{n}, i} ( \bm{q}_{\bm{n},i}^2 -1)^2 \ .
\end{equation}
The penalty parameter $\mu >0$ controls the trade-off between compactness and numerical stability.
For large but finite $\mu$, low-energy states remain confined near the compact submanifold and reproduce the same spectrum as the exact theory up to corrections of order $1/\mu$.

Both strategies are compatible with current CV technology. The first relies on experimentally achievable squeezing operations, while the second can be realized through engineered non-Gaussian CV gates.

As discussed in Section \ref{sec:3}, for a  square lattice with $(2N+1)\times (2N+1)$ plaquettes and open boundary conditions, there are $4(N+1)^2$ fermionic sites, and after gauge fixing, $(2N+1)^2$ independent gauge degrees of freedom.
Therefore, we need $\mathcal{O} (N^2)$ qubits and qumodes to realize the lattice system with DV/CV hybrid quantum hardware. Thus, the size of the quantum system needed scales linearly with the volume of the lattice system we wish to simulate. The gate complexity scales polynomially with lattice size when local interactions are decomposed via first- or second-order Trotterization. Truncating each qumode to 
$d_\text{eff}$ basis states yields a finite-dimensional approximation with total Hilbert-space dimension 
$2^{(2N+1)^2} d_\text{eff}^{4(N+1)(2N+1)}$, enabling classical benchmarking for small lattices and near-term quantum hardware implementations for few-plaquette systems.

\begin{figure}[t]
    \centering
    \includegraphics[width=0.45\linewidth]{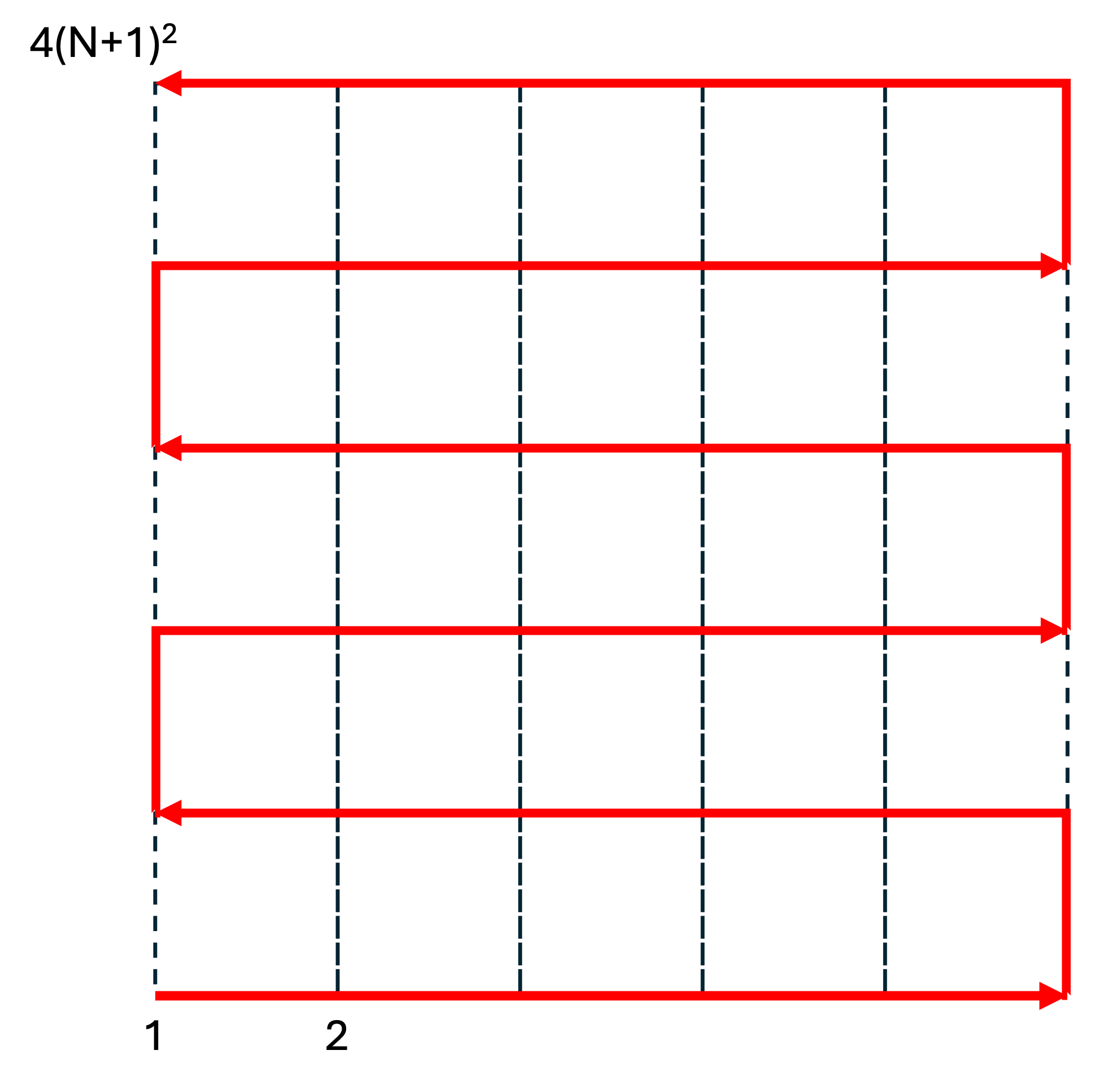}
    \caption{Path for the Jordan-Wigner transformation from fermions to spin degrees of freedom.}
    \label{fig:FermionicOrdering}
\end{figure}

Since the independent gauge degrees of freedom can be chosen along links in the $x$-direction, we will drop the index $x$ from these degrees of freedom in order to simplify the notation. Each can be simulated with a pair of qumodes $\bm{q}_{\bm{n}} = (q_{\bm{n}}^0, q_{\bm{n}}^1)$ with the electric field $E_{\bm{n}} \mapsto J_{\bm{n}} = q_{\bm{n}}^0 p_{\bm{n}}^1 - q_{\bm{n}}^1 p_{\bm{n}}^0$ and the link $U_{\bm{n}} \mapsto q_{\bm{n}}^0 +i q_{\bm{n}}^1$ (Eq.\ \eqref{eq:44}).
The various parts of the Hamiltonian (Eq.\ \eqref{eq:HEB}) are implemented as
\begin{eqnarray}\label{eq:411}
    H_E^{(Q)} &=& g^2\sum_{\bm{n},\bm{n}'} \left( \mathcal{H}_{\bm{n}\bm{n}'}^{(2)} J_{\bm{n}} J_{\bm{n}'} + \mathcal{H}_{\bm{n}\bm{n}'}^{(1)} Q_{\bm{n}} J_{\bm{n}'} + \mathcal{H}_{\bm{n}\bm{n}'}^{(0)} Q_{\bm{n}} Q_{\bm{n}'} \right) \ , \nonumber\\
    H_B^{(Q)} &=& \frac{1}{2g^2} \sum_{\bm{n}}   (\bm{q}_{\bm{n} + e_y} - \bm{q}_{\bm{n}})^2 \ , \nonumber\\
    H_M^{(Q)} &=& m_0 \sum_{\bm{n}} (-)^{n_x+n_y} Q_{\bm{n}} \ , \nonumber\\
    H_K^{(Q)} &=& \frac{1}{2} \sum_{\bm{n}}   (q_{\bm{n}}^0 - i q_{\bm{n}}^1 ) X_{\bm{n}}^+ X_{\bm{n}+ e_x}^- + \frac{1}{2}\sum_{\bm{n}}    P_{L(\bm{n}, \bm{n} + e_y)} X_{\bm{n}}^+  X_{\bm{n}+ e_y}^-+ \mathrm{h.c.} \ ,
\end{eqnarray}
where $P_{L(\bm{n}, \bm{n} + e_y)} = \prod_{\bm{n}'\in {L(\bm{n}, \bm{n} + e_y)}} Z_{\bm{n}'}$ is a string of Pauli $Z$ matrices along a path joining the sites $\bm{n}$ and $\bm{n} + e_y$.
Notice that once the constraint \eqref{eq:uc} is imposed, the magnetic Hamiltonian $H_B^{(Q)}$ reduces to the expected expression $H_B$. The form of $H_B^{(Q)}$ we chose ensures numerical stability. For the Jordan-Wigner transformation we chose the path illustrated in Figure \ref{fig:FermionicOrdering}. This is the most convenient choice that leads to simple two-qubits-one-qumode interactions in the kinetic Hamiltonian. Different paths could entail longer Pauli strings interacting with the bosonic modes.
The matrix $\mathcal{H}^{(2)}$ has diagonal entries ($\bm{n}=\bm{n}'$) equal to 2 and entries corresponding to nearest neighbors equal to $-1/2$.  $\mathcal{H}^{(1)}$ and $\mathcal{H}^{(0)}$ can be obtained by solving Gauss Law at each vertex and choosing an ordered path for both bosons and fermions, independently.  

\subsection{Time evolution}
To study the time evolution of the hybrid qubit-qumodes system, we need to engineer the incremental evolution operator by a small time interval $\Delta t$,
\begin{equation}
    e^{-i\Delta t H^{(Q)}} \approx e^{-i\Delta t H_E^{(Q)}} e^{-i\Delta t H_B^{(Q)}} e^{-i\Delta t H_M^{(Q)}} e^{-i\Delta t H_K^{(Q)}}.
    \label{eq:general_Trotter_HQ}
\end{equation}
\subsubsection*{Electric term}
The electric contribution in Eq.~\eqref{eq:general_Trotter_HQ} can be decomposed into factors of five different types,
\begin{equation}\label{eq:set}
    \left\{ e^{-isJ_{\bm{n}}} \ ,\ e^{-i s J^2_{\bm{n}}} \, \ e^{-isJ_{\bm{n}} J_{\bm{n}'}}\ , \ e^{-isZ_{\bm{n}} J_{\bm{n}'}} \ , \ e^{-isZ_{\bm{n}} } \ , \ e^{-isZ_{\bm{n}} Z_{\bm{n}'}} \right\},
\end{equation}
where $\bm{n}' \ne \bm{n}$. The first type of factor is a pure bosonic gate. Suppose that $J_{\bm{n}}$ acts on the pair of qubits with quadratures $\bm{q} = (q^0,q^1)$, so $J_{\bm{n}} =J^{01} = q^0 p^1 - q^1 p^0$. Using 
\begin{equation}
    J = {{\rm F}^0}^\dagger \cdot \text{BS} \cdot (N^0 - N^1) \cdot \text{BS} \cdot {\rm F}^0,
\end{equation}
where $N = \frac{1}{2} (p^2 + q^2)$ is the number operator, $F = e^{i\frac{\pi}{2} N}$ is the Fourier transform operator, and ${\text{BS}}$ represents a balanced beam splitter, we can express it as
\begin{equation}\label{eq:pure_Jterm}
    e^{-is J} = {\rm F^0}^\dagger \cdot \text{BS}^{\dagger} \cdot \text{R}^0 (-s) \cdot \text{R}^1 (s) \cdot \text{BS} \cdot {\rm F}^0.
\end{equation}
The quantum circuit implementing $e^{-is J}$ is
\[   \begin{quantikz}
        \lstick{$q^{0}$}\setwiretype{b} & \gate{\rm F} & \gate[2]{\text{BS}} & \gate{{\rm R}(-s)}  & \gate[2]{\text{BS}^{\dagger}} & \gate{\rm F^{\dagger}} & \rstick{} \\
        \lstick{$q^1$} \setwiretype{b}& & & \gate{{\rm R}(s)}&  & & \\
    \end{quantikz}
\]
Similarly, the second term in \eqref{eq:set} can be expressed as
\begin{equation}\label{eq:pure_J2term}
    e^{-is J^2} = {\rm F^0}^\dagger \cdot \text{BS}^{\dagger} \cdot \text{K}^0 (-s) \cdot \text{K}^1 (-s) \cdot \text{CK} (2s) \cdot \text{BS} \cdot {\rm F}^0
\end{equation}
in terms of single-qumode Kerr gates and two-qumode cross-Kerr gates (see Table \ref{table:1}). The quantum circuit implementing $e^{-is J^2}$ is
\[   \begin{quantikz}
        \lstick{$q^{0}$}\setwiretype{b} & \gate{\rm F} & \gate[2]{\text{BS}} & \gate{{\rm K}(-s)} & \gate[2]{{\rm CK}(2s)} & \gate[2]{\text{BS}^{\dagger}} & \gate{{\rm F}^{\dagger}} & \rstick{} \\
        \lstick{$q^1$} \setwiretype{b}& & & \gate{{\rm K}(-s)}&  & & & \\
    \end{quantikz}
\]
The second type of factor is also purely bosonic, but it contains two different $J$ operators acting on four qumodes. 
The quantum circuit implementing $e^{is J_{\bm{n}}\otimes J_{\bm{n}'}}$ on 2 gauge degrees of freedom is
\[ \begin{quantikz}[row sep =0.3cm]
        \lstick{$q^0_n$}\setwiretype{b} & \gate{\rm F} & \gate[2]{\rm{BS}} & & \gate[2]{\rotatebox{90}{${\rm CK}(s)$}} & & \permute{2,1} & & \permute{2,1} & & \gate[2]{\rm{BS}^{\dagger}} & \gate{\rm F^{\dagger}} & \\
        \lstick{$q^1_n$}\setwiretype{b} & &  & \permute{2,1} & & \gate[2]{\rotatebox{90}{${\rm CK}(-s)$}} & & \gate[2]{\rotatebox{90}{${\rm CK}(-s)$}} & & \permute{2,1} & & & \\
        \lstick{$q^0_{n'}$}\setwiretype{b} & \gate{\rm F} & \gate[2]{\rm{BS}} & & \gate[2]{\rotatebox{90}{${\rm CK}(s)$}} & & \permute{2,1} & & \permute{2,1} & & \gate[2]{\rm{BS}^{\dagger}} & \gate{\rm F^{\dagger}} &\\
        \lstick{$q^1_{n'}$}\setwiretype{b} &  &  & & & & & & & & & &\\
    \end{quantikz}
\]
The third type of factor, $e^{-is Z\otimes J}$, is a hybrid unitary involving two qumodes on which $J$ acts and one qubit on which $Z$ acts. It can be expressed as the hybrid qubit-qumode gate $\text{CBS} (z =- s)$ (see Table \ref{table:1}). 

The remaining factors in the set \eqref{eq:set} involve only Pauli matrices. They are implementable with standard qubit gates. 

\subsubsection*{Magnetic term}
The magnetic contribution in Eq.~\eqref{eq:general_Trotter_HQ} can be decomposed into two different sets of purely bosonic contributions
\begin{equation}
    \left\{ e^{-is\hat q^2}, e^{-is\hat q}, e^{-is (\hat q_1 - \hat q_2)^2} \right\}.
\end{equation}
They are all Gaussian qumode gates that can be implemented with quadratic phase gates, displacements, and balanced beam splitters (see Table \ref{table:1}).

\subsubsection*{Kinetic term}
The fermionic kinetic contribution contains purely qubit gates as well as hybrid gates of the form (see Eq.\ \eqref{eq:411})
\begin{equation}
    e^{i s \hat{q}\otimes \sigma_i\otimes \sigma_j},
\end{equation}
They can be implemented with a doubly controlled displacement targeting the qumode with qubits acting as control operators,
\begin{equation}
   \text{CCD} (s) = e^{-is\hat{q}\otimes Z\otimes Z} = \left(\mathbb{1}\otimes\text{CNOT}\right)\cdot \left( e^{-is\hat{q}\otimes Z} \otimes \mathbb{1} \right) \cdot \left(\mathbb{1}\otimes\text{CNOT} \right),
\end{equation}
where we used the identity
$Z\otimes Z = \text{CNOT} \cdot (Z\otimes \mathbb{1})\cdot \text{CNOT}$.
In order to obtain the desired qubit controls we act with appropriate single qubit rotations.
For example, for the unitary operator $e^{-i s\hat{q}\otimes X\otimes X}$, we obtain
\begin{equation}\label{eq:428}
    e^{-i s\hat{q}\otimes X\otimes X} = (H\otimes H) \cdot \text{CCD} (s) \cdot (H\otimes H),
\end{equation}
where $H$ is the Hadamard matrix.
The quantum circuit implementing the unitary evolution operator \eqref{eq:428} is
\begin{equation*}
  e^{-is \hat{q}\otimes X\otimes X} = \begin{quantikz}
\setwiretype{b}\qw &       &  & \gate{D(s)}  & \qw      & \qw      & \rstick{qumode} \\
 &    \gate{H} & \targ{} & \ctrl{-1}  &   \targ{} &\gate{H}    &       & \rstick{qubit }\\
  & \gate{H}  &  \ctrl{-1}  &      & \ctrl{-1} &\gate{H} &      & \rstick{qubit}
\end{quantikz}
\end{equation*}
The unitary evolution associated with the other terms in $H_K^{(Q)}$ in Eq.~\eqref{eq:411} is implementable with standard qubit gates.

\subsubsection*{Mass term}
Finally, the fermion mass term in the Hamiltonian \eqref{eq:411} is implementable with single qubit $Z$ rotations.

\subsection{QITE}\label{sec:QITE}

One way to obtain the ground state of any given theory is to exploit the rapid decay of the evolution operator evaluated at imaginary times \cite{Motta2020,PhysRevA.105.012412},
\begin{equation}\label{eq:general_Qite}
    \frac{\bra{\psi}e^{-2\tau H} H\ket{\psi}}{\bra{\psi}e^{-2\tau H}\ket{\psi}}= E_0 + \mathcal{O}(e^{-2\tau (E_1-E_0)}),
\end{equation}
where $\ket{\psi}$ is an arbitrary state with non-vanishing overlap with the ground state.

Unlike standard qubit-based QITE algorithms, the present formulation operates on a hybrid Hilbert space incorporating both discrete (qubit) and continuous (qumode) generators. Each imaginary-time update simultaneously adjusts matter amplitudes and gauge-field quadrature distributions while preserving the Gauss-law constraints. To our knowledge, this represents the first implementation of QITE in a hybrid DV–CV gauge-theory setting.

Rotating to the imaginary axis is a non-unitary procedure that necessarily involves a measurement. We introduce two different methods to engineer the necessary gates for QITE.

\subsubsection*{QITE Method A}

For a Trotter step, we need to engineer the non-unitary operator $e^{-\Delta \tau \hat{O}}$, where $\hat{O}$ is a Hermitian hybrid operator, e.g., $\hat{O} = \hat{h}\otimes P$. Here, $\hat{h}$ acts on a register of qumodes and $P$ is a string of Pauli operators acting on qubits.

We introduce an ancilla qubit in the state $\ket{0}_a$ and act with the entangling unitary operator $e^{i\Delta \tau \hat{O} \otimes Y_a}$, where $Y_a$ is a Pauli matrix acting on the ancilla qubit, followed by a Hadamard gate $H_a$. After measuring the ancilla projecting it onto the $\ket{0}_a$ state, we obtain the operator
\begin{equation}
    {}_a \bra{0} H_a e^{i\Delta \tau \hat{O} \otimes Y_a} \ket{0}_a = \frac{1}{\sqrt{2}}e^{-\Delta \tau \hat{O}} + \mathcal{O} ((\Delta \tau)^2)\,,
\end{equation}
The quantum circuit that implements this is
\begin{equation*}
  \begin{quantikz}
\lstick{qumodes}\setwiretype{b}  &   & \gate[wires=3]{e^{i\Delta \tau \hat{O} \otimes Y}} &  & &  \\
 \lstick{qubits}&  & &  & & \\
 \ket{0}_a &   &  & \gate{H}&\meter{} & \arrow[r] & \ket{0}_a 
\end{quantikz}
\end{equation*}
By repeatedly applying this circuit, we engineer the non-unitary $e^{-\tau O}$ for finite imaginary time $\tau$.

\subsubsection*{QITE Method B}

Alternatively, the non-unitary $e^{-\Delta \tau \hat{O}}$ can be engineered with the aid of an ancilla qumode. An advantage of this method is that $\Delta \tau$ does not need to be small. However, it can only be applied to Hermitian operators $O$ that can be brought into the form $\hat{O} \to \hat{o}^2$, where $\hat{o}$ is a hybrid Hermitian operator. Assuming that this is the case, i.e., 
\begin{equation}\label{eq:U}
    \hat{O} = U^\dagger \hat{o}^2 U
\end{equation}
where $U$ is a unitary operator, 
we introduce an ancilla qumode in the vacuum state $\ket{0}_A$ \cite{PhysRevA.105.012412}, and apply the entangling non-Gaussian unitary gate
\begin{equation}
 e^{i 2\sqrt{\Delta\tau} \hat{o} \otimes \hat{p}_A}   ,
\end{equation}
where $\hat{p}_A$ is a quadrature operator acting on the ancilla qumode.
We then measure the ancilla qumode with a photon detector. If the detector detects no photon, the measurement acts as a projection onto
\begin{equation}
\begin{split}
    &{}_A \bra{0} e^{i 2\sqrt{\Delta\tau} \hat{o} \otimes \hat{p}_A} \ket{0}_A \propto e^{-\Delta\tau \hat{o}^2}.
\end{split} 
\end{equation}
After applying the unitary transformation \eqref{eq:U}, we obtain the desired non-unitary operator,
\begin{equation}
    U^\dagger e^{-\Delta\tau \hat{o}^2} U = e^{-\Delta \tau \hat{O}} .
\end{equation}
The quantum circuit that implements this is 
\begin{equation*}
  \begin{quantikz}
\lstick{qumodes}\setwiretype{b}  & \gate[wires=2]{U}   & \gate[wires=3]{e^{i2\sqrt{\Delta \tau} \hat{o}\otimes \hat{p}}} & \gate[wires=2]{U^\dagger} &  \\
 \lstick{qubits} & &  &  &  \\
 \ket{0}_A \setwiretype{b} & &  &  & \meterD{n=0}  
\end{quantikz}
\end{equation*}
For example, consider $\hat{O} = 2g^2 J^2$ which contributes to the electric part of the Hamiltonian. In this case, $\hat{o} = \sqrt{2} g J$ and $U = \mathbb{1}$. 

The hybrid framework developed in this Section provides general algorithms for the calculation of physical properties, such as the QITE algorithm for preparing ground and low-lying excited states of lattice gauge theories encoded on qubit–qumode architectures. To illustrate its implementation and to benchmark convergence and gauge-invariance preservation, we apply our method to the minimal non-trivial instance of 
U(1) quantum electrodynamics in 
2+1 dimensions, the single-plaquette model, in Section \ref{sec:single}. This system contains the full structure of the theory, including compact electric and magnetic terms, local Gauss-law constraints, and fermionic matter coupling, yet remains analytically tractable for comparison with exact diagonalization. The single-plaquette problem thus serves as a testbed for validating the hybrid encoding, assessing the effectiveness of the squeezing- and Hamiltonian-penalty compactness schemes, and quantifying the accuracy of hybrid algorithms in reproducing gauge-invariant dynamics.

\section{Results for a single plaquette}
\label{sec:single}

To illustrate our hybrid quantum approach introduced in Section \ref{sec:4}, in this Section we derive explicit results for a lattice consisting of a single plaquette and compare them with analytic expressions and exact diagonalization (ED).

\subsection{The Hamiltonian}
In our hybrid quantum computation, we implement the gauge field with a pair of qumodes of quadratures $\bm{q} = (q^0,q^1)$ and $\bm{p} = (p^0,p^1)$ obeying the commutation rules $[q^\mu, p^\nu] = i\delta^{\mu\nu}$, and the fermionic field with four qubits. After imposing three of the Gauss Law constraints, the remaining variable $E$ is implemented as an angular momentum $J = q^0p^1 - q^1 p^0$,
with the corresponding link variable implemented as $U = q^0 + iq^1$. This enlarges the phase space, but it does not break gauge invariance because the angular momentum $J$ leaves the unit circle $\bm{q}^2 =1$ invariant, $[J, \bm{q}^2] =0$. As a consequence, $\bm{ q}^2$ is conserved, i.e. $[H,\bm{q}^2]=0$. 

For the fermionic field, we employ the Jordan-Wigner transformation, implementing it via Pauli matrices,
\begin{equation}
        \Psi_1\mapsto X_1^-,\ \
        \Psi_2\mapsto -iZ_1X_2^-,\ \
        \Psi_3\mapsto -Z_1 Z_2 X_3^-,\ \
        \Psi_4\mapsto iZ_1 Z_2 Z_3 X_4^- , 
\end{equation}
where $X^\pm = \frac{1}{2} (X \pm iY)$ and the indices indicate the position of the qubit the Pauli matrix acts on. In this representation, the charge operators become projection operators acting on single qubits,
\begin{equation}
        Q_1 \mapsto  \frac{Z_1 + \mathbb{1}}{2}\ ,\ \
        Q_2 \mapsto  \frac{Z_2 - \mathbb{1}}{2}\ ,\ \
        Q_3 \mapsto  \frac{Z_3 + \mathbb{1}}{2}\ ,\ \
        Q_4 \mapsto \frac{Z_4 - \mathbb{1}}{2}\ .
\end{equation}
We model the system with the Hamiltonian
\begin{equation}\label{eq:HQ0}
    H^{(Q)} = H_E^{(Q)} + H_B^{(Q)} + H_M^{(Q)} + H_K^{(Q)}
\end{equation}
where
\begin{equation}\label{eq:HQ1}
    \begin{split}
        H_E^{(Q)} &= \frac{g^2}{2}\left[J^2 + \left(J-Q_2\right)^2 + \left(J+Q_3\right)^2 + \left(J - Q_1 - Q_2\right)^2\right] , \\
        H_B^{(Q)} &= \frac{1}{2g^2} [ (q^0-1)^2 + (q^1)^2 ]  , \\
        H_M^{(Q)} & = m_0\sum_{i=1}^4(-1)^{1+i} Q_i,\\
    H_K^{(Q)} & = \frac{1}{2} \left(X^+_1 X_2 + (q^0 - iq^1) X^+_3 X_2^- + X^+_3 X^-_4 + X^+_1 Z_2 Z_3 X^-_4  \right) + \text{h.c.}\ .
    \end{split}
\end{equation}
and $J = q^0 p^1 - q^1 p^0$.

The remaining Gauss's Law constraint \eqref{eq:qt} imposing vanishing total charge will be implemented by only considering states obeying \eqref{eq:qt}. This restricts the 16-dimensional fermionic Hilbert space to a six-dimensional subspace. Thus, it is possible to describe the system with three qubits. However, the representation of fermionic operators becomes cumbersome, resulting in a longer quantum circuit. In the following, we describe the system with four qubits.

\begin{figure}[t]
     \centering
 \begin{quantikz}
     \lstick{$\ket{0}_a$}\setwiretype{b} & \gate{D(is)} & \gate{S(r)} & & \gate{\mathcal{D}(s)} & \gate{\mathcal{D}(s)} & \meterD{p=0}\\
     \lstick{$\ket{0}_0$}\setwiretype{b} &  \gate{D(\alpha)} & & &\ctrl{-1} & &  \\
     \lstick{$\ket{0}_1$}\setwiretype{b} &  & &  & & \ctrl{-2} & 
 \end{quantikz}
     \caption{Quantum circuit that creates the ground state coupled cluster ansatz $\ket{\psi_0 (\alpha, \Lambda)}$, see Eq.\ \eqref{eq:GroundState_ansatz}, with radial cutoff $\Lambda = \frac{g}{2} se^r$.}
     \label{circ:2}
 \end{figure}
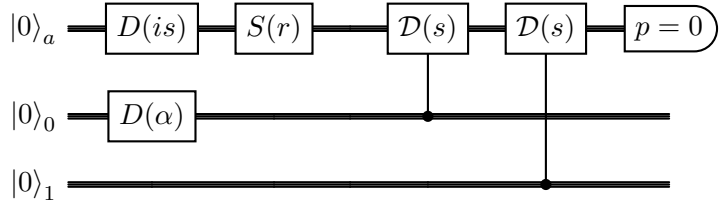

\subsection{Variational ansatz}

We consider the pure U(1) gauge theory and the ground state ansatz 
\begin{equation}
         \ket{\psi_0(\alpha,\Lambda)} \propto \int d^2q\,e^{-\frac{\Lambda^2}{g^2} \left(\bm{q}^2 -1\right)^2} e^{\alpha q^0}e^{-\frac{1}{2} \bm{q}^2}\ket{\bm{q}}\ ,
    \label{eq:GroundState_ansatz}
\end{equation}
which can be engineered with the quantum circuit depicted in Figure \ref{circ:2} with the gate parameters $r,s$ chosen so that $\Lambda = \frac{g}{2} s e^r$ (Eq.\ \eqref{eq:82m}). Without fermions, the Hamiltonian operator is given by
\begin{equation}
    H_g^{(Q)} = H_E^{(Q)} + H_M^{(Q)} \ , 
\end{equation}
with vanishing charges, $Q_i=0$ ($i=1,2,3,4$). We obtain the energy of the ansatz \eqref{eq:GroundState_ansatz} in terms of an integral over the unphysical radial coordinate $R = \sqrt{\bm{q}^2}$ and Bessel functions we obtain by integrating of the angular variable $\chi = \arctan \frac{q^1}{q^0}$,
\begin{equation}
    \begin{split}
        \epsilon_0(\alpha,\Lambda) &\equiv \bra{\psi_0(\alpha,\Lambda)}  H_g^{(Q)} \ket{\psi_0(\alpha,\Lambda)} \\
        & = \frac{ \int_0^{\infty}dR\,R\,   e^{-2\frac{\Lambda^2}{g^2} \left(R^2-1\right)^2 -R^2} \left(  \left(R^2+1\right) I_0(2 R \alpha )+2 R (g^4\alpha -1 ) I_1(2 R \alpha )\right)}{2g^2 \int_0^{\infty}dR\,R\, e^{-2 \frac{\Lambda^2}{g^2} \left(R^2-1\right)^2 -R^2} I_0(2 R \alpha )}.
    \end{split}
\end{equation}
These integrals can be evaluated for large radial cutoff $\Lambda$
by expanding around $R=1$, 
\begin{figure}[t]
    \centering
    \includegraphics[width=0.6\linewidth]{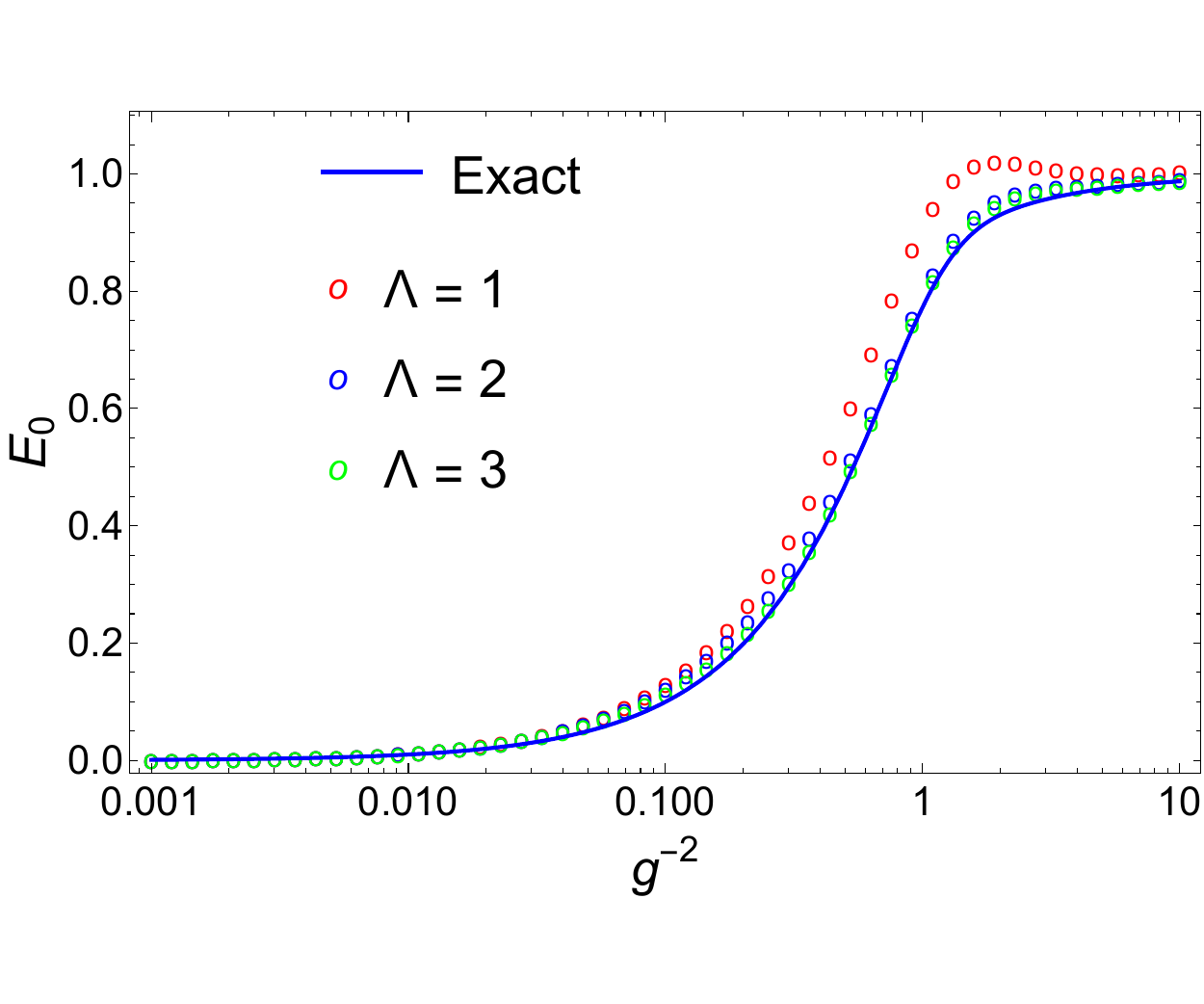}
    \caption{Ground state energy of a single-plaquette pure-gauge system \emph{vs.}\ the coupling constant using a variational method for various values of the radial cutoff $\Lambda$.}
    \label{fig:E0squeezing_Lambdas}
\end{figure}
and using
\begin{equation}\label{eq:510} \int_0^\infty dR\, R\, e^{-2\frac{\Lambda^2}{g^2} (R^2 -1)^2} (R^2 -1)^{N} = \left\{ \begin{array}{cc}
   \sqrt{\frac{\pi}{2}} \frac{(N-1)!!}{(2\Lambda/g)^{N+1}} + \mathcal{O} (e^{-2\Lambda^2/g^2}) & , \ \ N\ \text{even}  \\
   \mathcal{O} (e^{-2\Lambda^2/g^2})  & , \ \ N\ \text{odd}
\end{array} \right.  \end{equation}
We obtain the expansion
\begin{equation}
    \epsilon_0 (\alpha, \Lambda) = \frac{1}{g^2   } \left( 1  - \frac{I_1 (2\alpha)}{I_0 (2\alpha) } \right) + g^2\alpha  \frac{I_1 (2\alpha) }{I_0 (2\alpha)} + \mathcal{O} \left( \frac{1}{\Lambda^2} \right).
\end{equation}
In the weak coupling limit, it is minimized for
$\alpha \approx \frac{1}{2g^2}$.
We deduce the estimate of the ground state energy
\begin{equation}\label{eq:E0L}
    E_0 = 1 + \frac{1}{8\Lambda^2}  + \mathcal{O} (g^2, \Lambda^{-4}).
\end{equation}
Figure \ref{fig:E0squeezing_Lambdas} shows results obtained by minimizing the energy of the ansatz over the variational parameter $\alpha$ for various values of the radial cutoff ($\Lambda = 1,2,3$) as a function of the coupling constant $g$.
Very good agreement with exact results is  found even for small values of $\Lambda$. Ground state energy levels asymptote to $E_0 \to 1$ in the weak coupling limit ($g\to 0$), as expected from the analytical result \eqref{eq:E0L}.

\begin{figure}[t]
    \centering
    \includegraphics[width=0.45\linewidth]{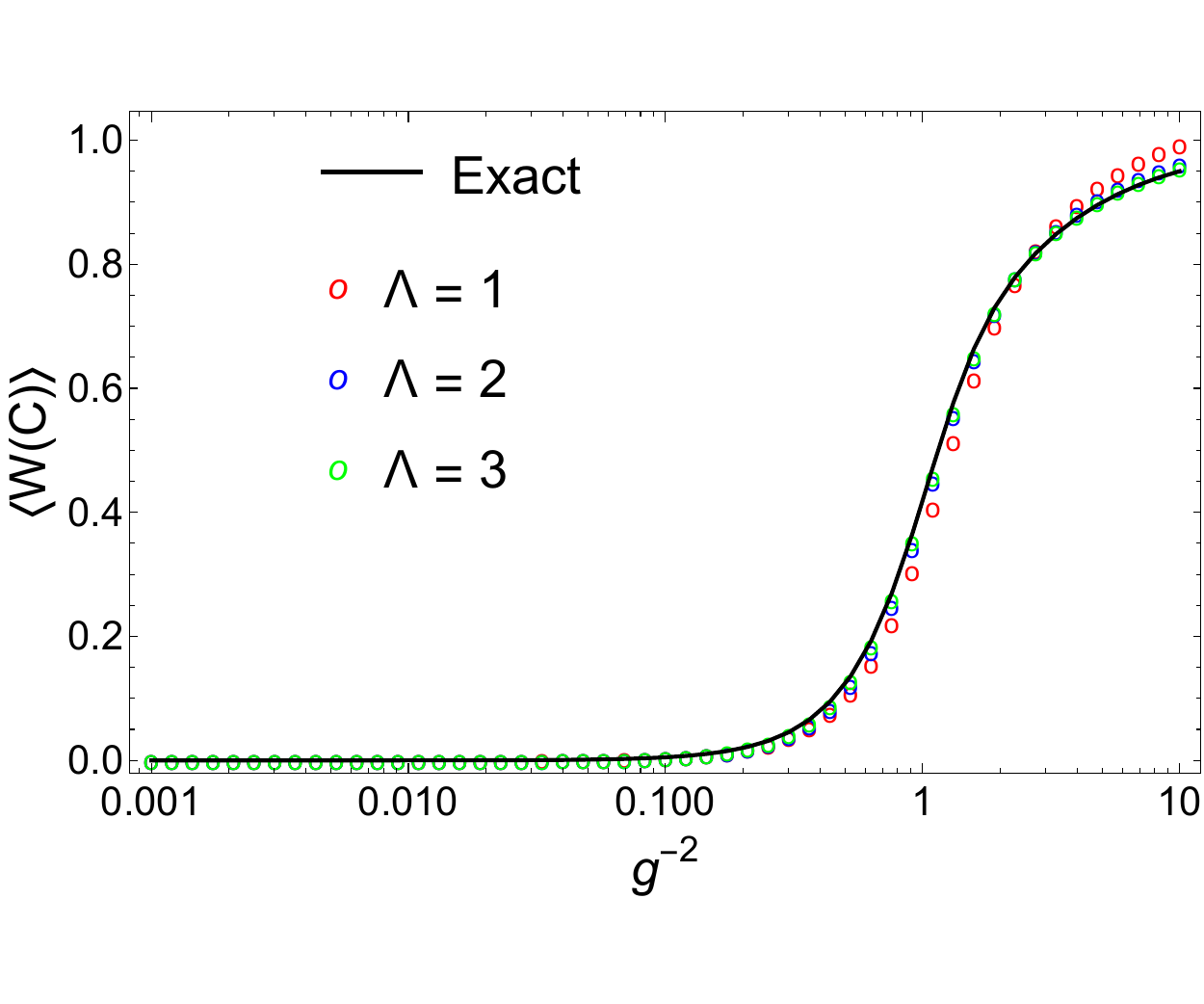}
\quad\includegraphics[width=0.45\linewidth]{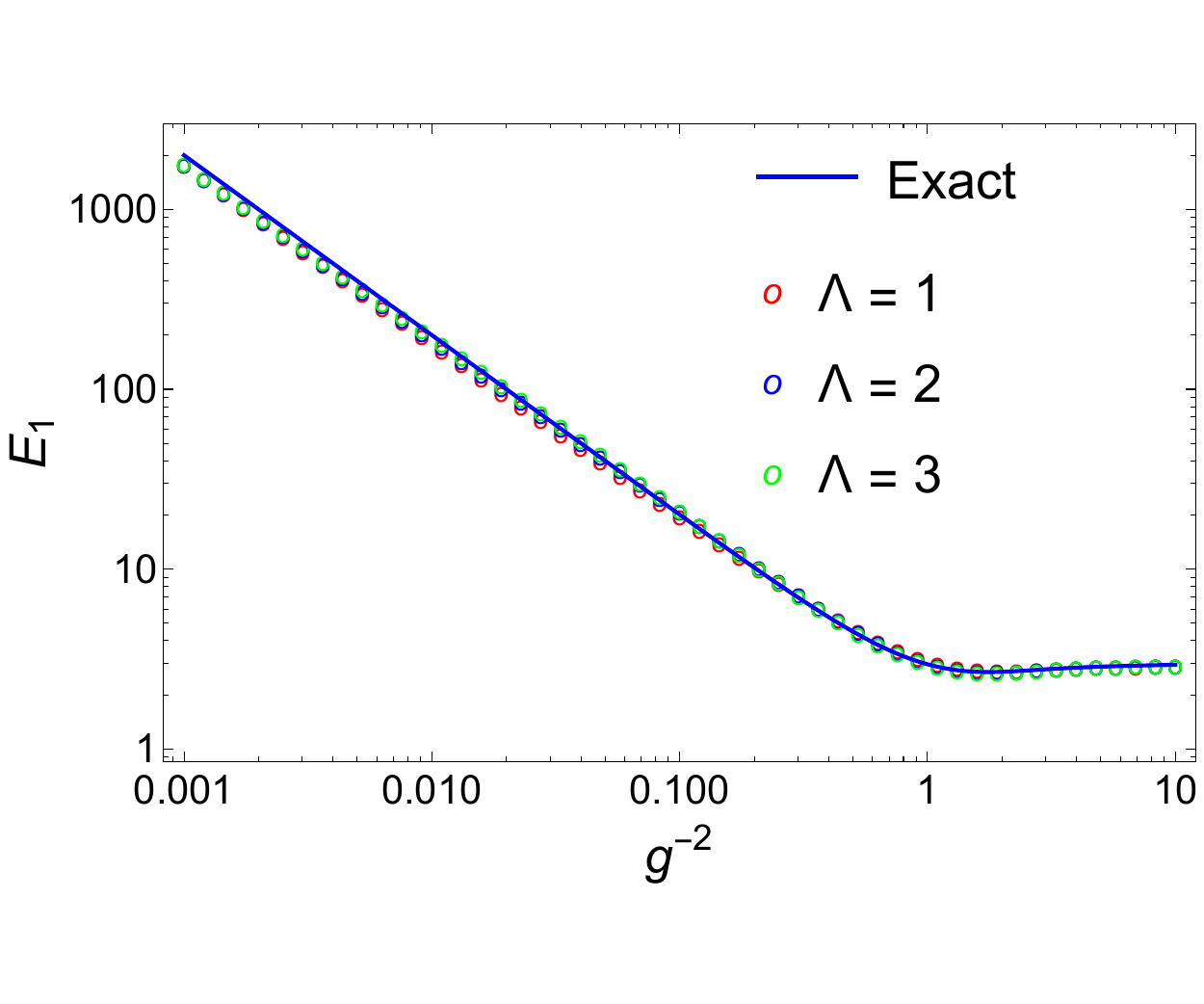}
       \caption{Vacuum expectation value of the plaquette operator (left) and first photonic energy level (right) of a single-plaquette pure-gauge system \emph{vs.}\ the coupling constant using a variational method for various values of the radial cutoff $\Lambda$.  }
    \label{fig:PlaqOp_groundstate_exp}
\end{figure}
Similarly, the expectation value of the constraint ($\bm{q}^2 =1$) can be shown to vanish as $\Lambda\to\infty$,
\begin{equation}
    \begin{split}
        \bra{\psi_0(\alpha,\Lambda)}  (\bm{q}^2-1)^2\ket{\psi_0(\alpha,\Lambda)} & =\frac{\int_0^{\infty}dR\,R\,  \left(R^2-1\right)^2 e^{-2\frac{\Lambda^2}{g^2} \left(R^2-1\right)^2 -R^2} I_0(2 R \alpha )}{ \int_0^{\infty}dR\,R\, e^{-2\frac{\Lambda^2}{g^2} \left(R^2-1\right)^2 -R^2} I_0(2 R \alpha )} \\
        & = \frac{g^2}{4\Lambda^2} + \mathcal{O} \left( \frac{1}{\Lambda^4} \right) \ ,
    \end{split}
\end{equation}
as desired. Using the ground state ansatz \eqref{eq:GroundState_ansatz}, we obtain the estimate of the vacuum expectation value of the plaquette operator,
\begin{equation}
    \begin{split}
\braket{W[C]} = \bra{\psi_0(\alpha,\Lambda)}q^0\ket{\psi_0(\alpha,\Lambda)} & = \frac{\int_0^{\infty}\text{d}R\,R^2\, e^{-2\frac{\Lambda^2}{g^2} \left(R^2-1\right)^2 -R^2} I_1(2 R \alpha )}{\int_0^{\infty}\text{d}R\,R\, e^{-2\frac{\Lambda^2}{g^2} \left(R^2-1\right)^2 -R^2} I_0(2 R \alpha )}\ .
    \end{split}
\end{equation}
Using Eq.\ \eqref{eq:510}, we obtain the expansion
\begin{equation}
    \braket{W[C]} = \frac{I_1 (2\alpha)}{I_0 (2\alpha)} + \frac{g^2 \alpha}{8\Lambda^2} \left( 3 \left( \frac{I_1 (2\alpha)}{I_0 (2\alpha)} \right)^2 -2 \right) + \mathcal{O} \left(  \frac{1}{\Lambda^4} \right) \ .
\end{equation}
In particular, in the weak coupling limit, we have $\alpha \approx \frac{1}{2g^2}$, therefore,
\begin{equation}\label{eq:WCas}
    \braket{W[C]} = 1 - \frac{g^2}{2}  + \frac{1}{16\Lambda^2} \left( 1 - {3} g^2  \right) + \mathcal{O} (g^4, \Lambda^{-4}) \ .
\end{equation}
Similarly, using the ansatz for the first excited state, 
\begin{equation}
    \ket{\psi_1(\beta,\Lambda)} = q^1\ket{\psi_0(\beta,\Lambda)},
    \label{eq:FirstExcitedState_ansatz}
\end{equation}
we obtain the estimate of the first photonic energy level,
\begin{equation}
    \begin{split}
         \epsilon_1 &\equiv \frac{\bra{\psi(\beta,\Lambda)}  H \ket{\psi_1(\beta,\Lambda)}}{\bra{\psi_1(\beta,\Lambda)}\ket{\psi_1(\beta,\Lambda)}}\\
         & = \frac{\int_0^{\infty}\text{d}R\,R^2\, e^{-2\frac{\Lambda^2}{g^2} \left(R^2-1\right)^2-R^2} \left( \left(2  +\beta  \left( 1 + R^2 -2g^4\right) \right) I_1(2R\beta) -2\beta (1-3g^4 \beta   )R\, I_0(2 R \beta)\right)}{2\beta g^2\int_0^{\infty}\text{d}R\,R^2\, e^{-2\frac{\Lambda^2}{g^2} \left(R^2-1\right)^2-R^2} I_1(2 R \beta )} \\
         & =     \frac{1}{g^2} \left( 1 + \frac{1}{ \beta } -g^4\right) - \left( \frac{1}{g^2} - 3g^2 \beta   \right) \frac{  I_0(2  \beta)}{ I_1(2  \beta )} + \mathcal{O} \left(  \frac{1}{\Lambda^2} \right) \ .
    \end{split}
\end{equation}
In the weak coupling limit, this is minimized for $\beta \approx \frac{1}{2g^2}$. We deduce the estimate of the first photonic energy level and the gap, 
\begin{equation}\label{eq:E1}
    E_1 = 3 + \frac{1}{8\Lambda^2} + \mathcal{O} (g^2, \Lambda^{-4}) \ , \ \ \Delta E = E_1 - E_0 = 2 + \mathcal{O} (g^2, \Lambda^{-4}).
\end{equation}
We obtain very good agreement with exact results for both the vacuum expectation value of the Wilson loop $\braket{W[C]}$ and the first photonic energy level, as shown in Figure \ref{fig:PlaqOp_groundstate_exp}. In the weak coupling limit ($g\to 0$), the graphs asymptote to the values obtained by analytical calculations ($W[C] \to 1$, in agreement with Eq.\ \eqref{eq:WCas}, and $E_1 \to 3$, in agreement with Eq.\ \eqref{eq:E1}).

\begin{figure}[t]
    \centering
    \includegraphics[width=0.6\linewidth]{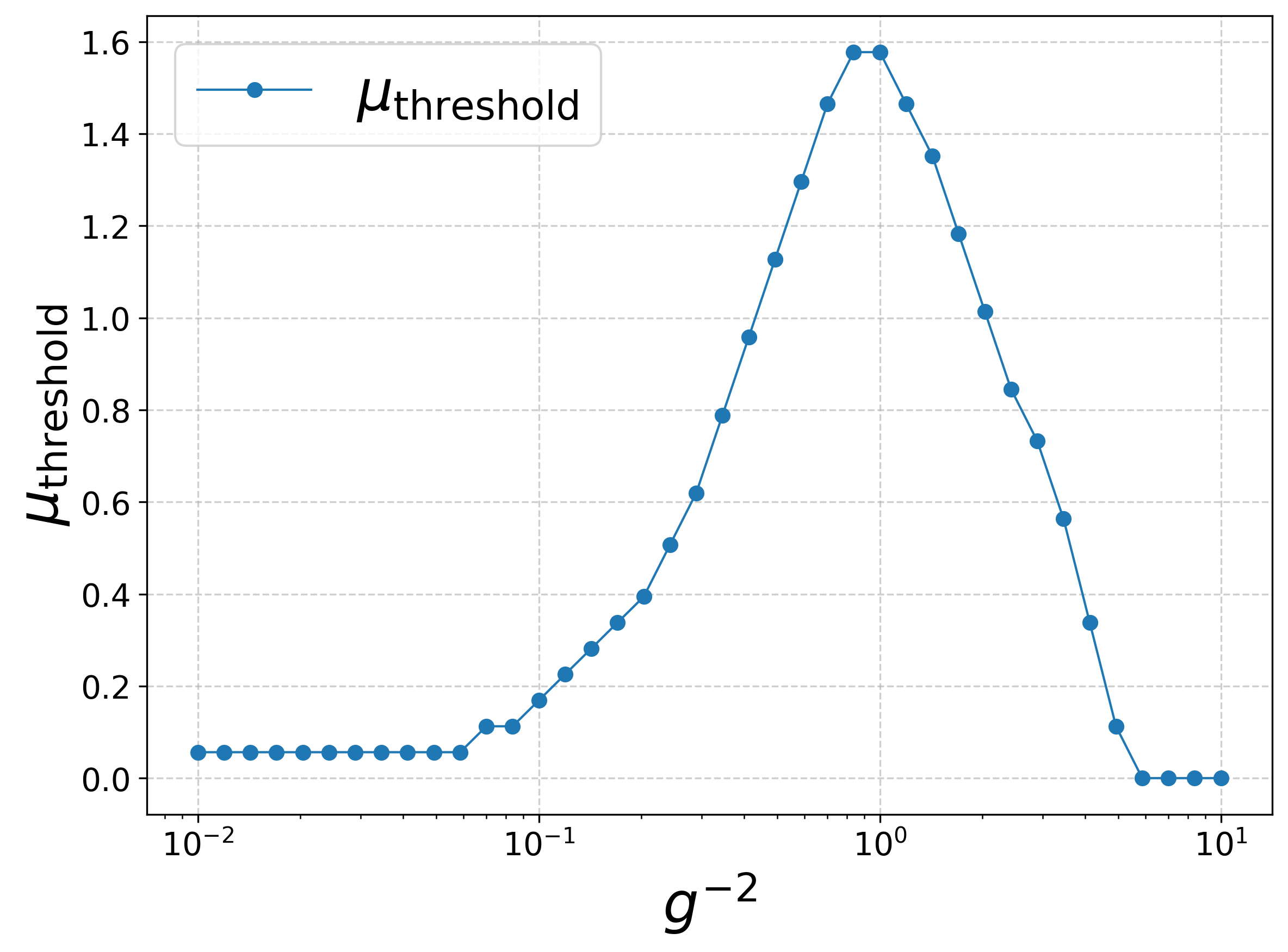}
    \caption{The threshold value of $\mu$ for an acceptable penalty term \eqref{eq:518} in the Hamiltonian of a single plaquette \emph{vs.}\ the coupling constant. Fermion mass is set to $m_0 = 1.5$.}
    \label{fig:mu_threshold_vs_gm2}
\end{figure}

\subsection{Hamiltonian penalty method}

As discussed in Section \ref{sec:4}, a second approach to enforcing the compactification constraint is by adding a penalty term to the Hamiltonian (Eq.\ \eqref{eq:LM}),
\begin{equation}\label{eq:518}
    H_\mu = \frac{\mu}{2}\left(\hat{q}_0^2+\hat{q}_1^2-1\right)^2 = \frac{\mu}{2}(R^2-1)^2 \quad (\mu>0)\ .
\end{equation}
Clearly, for sufficiently large $\mu$, the ground state energy corresponds to a configuration for which $R =1$. In order to illustrate the mechanism, we compute the ground state by exactly diagonalizing the Hamiltonian. Since the (unphysical) radial coordinate $R$ is non-dynamical, it effectively behaves as a parameter of the theory. The case with $\mu = 0$ corresponds to no constraint being imposed, and the energy corresponding to the physical configuration is not the one with minimal energy. As we increase $\mu$ above a certain value, the ground state favors the physical choice $R = 1$. The threshold value of $\mu$ for an acceptable penalty term \eqref{eq:518} in the Hamiltonian of a single plaquette \emph{vs.}\ the coupling constant is shown in Figure \ref{fig:mu_threshold_vs_gm2}. We set the fermion mass $m_0 =1.5$. Similar results are obtained for different choices of the fermion mass parameter. The threshold value was chosen for ground state energies within 1\% of the exact ground state energy and radial coordinate within 5\% of the physical value $R=1$. Interestingly, for all values of the coupling constant, a choice of $\mu > 1.6$ suffices. Moreover, in the weak coupling limit, the penalty term is not needed as the physical choice $R=1$ is naturally favored.

\begin{figure}[t]
    \centering
    \includegraphics[width=0.6\linewidth]{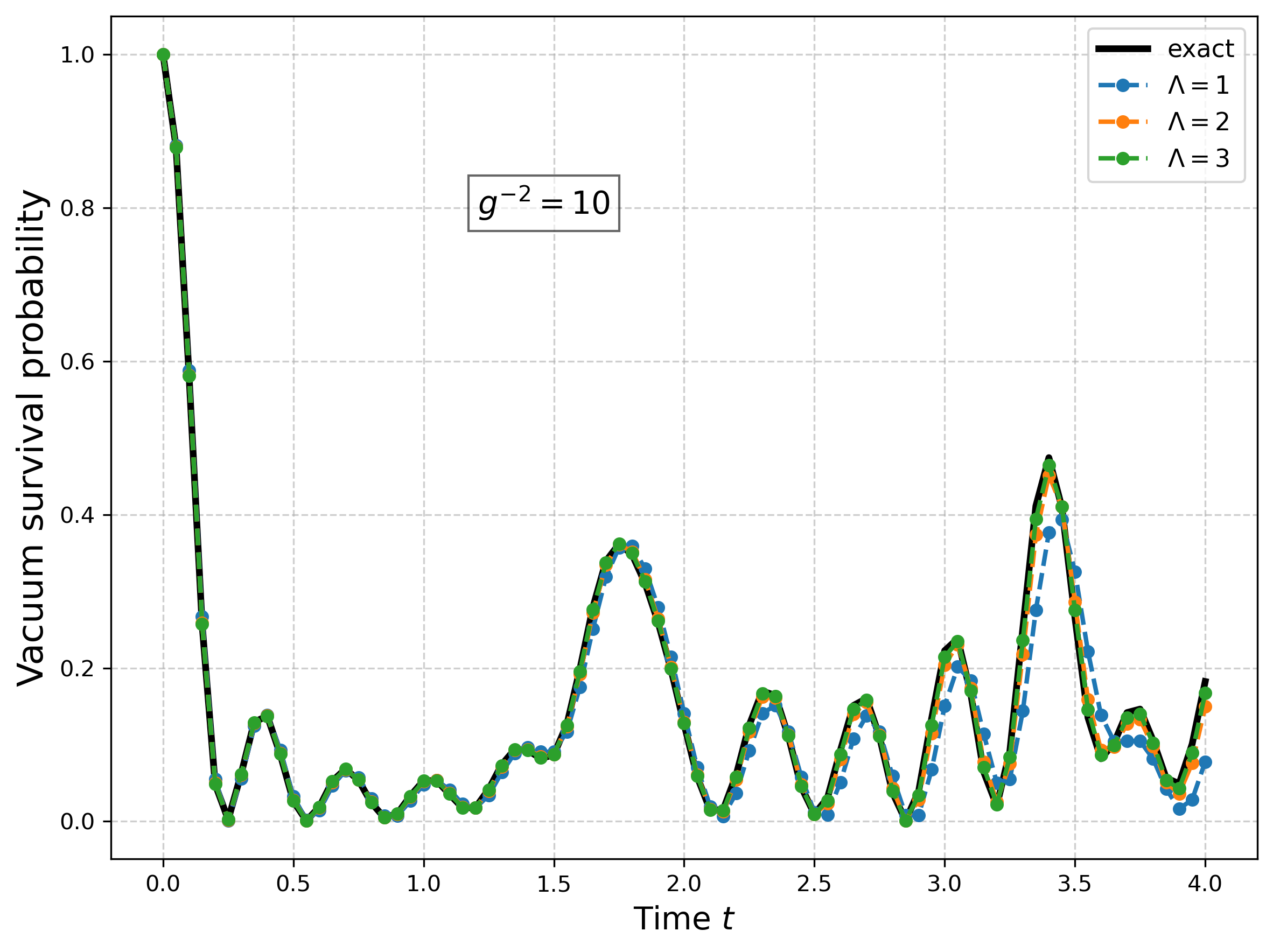}
    \caption{Survival probability of the state $\ket{V}$,  see Eq.\ \eqref{eq:free_vacuum_uc}, as a function of time for different values of the squeezing parameter $\Lambda$ compared with the exact results for a single plaquette.}
    \label{fig:vacuum_survival_gm2}
\end{figure}

\subsection{Transition probabilities}

Next, we consider the time evolution of the complete single-plaquette system including fermions, and perform the quantum computation of transition probabilities using the quantum circuits introduced in Section \ref{sec:4}.
As an example, we will work with the initial state
\begin{equation}
    \ket{V}\equiv\ket{vvvv}\otimes \ket{\psi_0(\alpha =0,\Lambda)}\ .
    \label{eq:free_vacuum_uc}
\end{equation}
Here, $\ket{vvvv}$ is the ground state of free static fermions, Eq.\ \eqref{eq:wff}, and $\ket{\psi_0}$ is in the gauge sector and defined in Eq.\ \eqref{eq:GroundState_ansatz}. We compute the survival probability of the state $\ket{V}$ under time evolution with Hamiltonian $H^{(Q)}$, Eq.\ \eqref{eq:HQ0}, with its components given by Eq.\ \eqref{eq:HQ1}. We Trotterize the evolution operator as
\begin{equation}\label{eq:5T}
    e^{-it H^{(Q)}} = \left( e^{-i\Delta t H_E^{(Q)}} e^{-i\Delta t H_B^{(Q)}}e^{-i\Delta t H_M^{(Q)}}e^{-i\Delta t H_K^{(Q)}} \right)^{t/\Delta t}  .
\end{equation}
The survival probability is obtained by applying the string of gates in \eqref{eq:5T} to the state $\ket{V}$ and then measuring the qubits and qumodes projecting back onto the state $\ket{V}$. The quantum circuit implementing this is 
\[  
\begin{quantikz}
     \lstick{$\ket{0}_a$}\setwiretype{b} & \gate{D(is)} & \gate{S(r)} & & \gate{\mathcal{D}(s)} & \gate{\mathcal{D}(s)} & & & \meterD{p=0}\\
     \lstick{$\ket{0}_0$}\setwiretype{b} &   & & &\ctrl{-1} & & \gate[wires=3]{e^{-it H^{(Q)}}}& & \meterD{n=0}\\
     \lstick{$\ket{0}_1$}\setwiretype{b} &  & &  & & \ctrl{-2} & & & \meterD{n=0}\\
     \lstick{$\ket{vvvv}$} &  & &  & &  & & \meter{} & \arrow[r] & \ket{vvvv}
 \end{quantikz}
\]
To simulate the survival amplitude, we use polar coordinates for the qumodes ($q^0 = R \cos\chi$, $q^1 = R \sin\chi$) and
apply ED (exact diagonalization) in the gauge field basis $\braket{\chi \vert j}= \frac{1}{\sqrt{2\pi}} e^{ij\chi}$, $j\in\mathbb{Z}$, where $J\ket{j} = j\ket{j}$, and the fermionic field basis $\ket{v_i}$ ($i=1,2,\dots, 6$) defined in Eq.\ \eqref{eq:wff}. We obtain the following expression for the complex amplitude
\begin{equation}\label{eq:520}
\begin{split}
    \mathcal{A} (t) &\equiv \langle V|e^{-itH^{(Q)}}|V\rangle \\ & = \frac{\int_0^{\infty}\text{d}R\,R\,\,e^{-R^2}e^{-2\frac{\Lambda^2}{g^2} (R^2-1)^2}\bra{j=0}\otimes\bra{vvvv}e^{-it H^{(Q)}}\ket{vvvv}\otimes\ket{j=0}}{\int_0^{\infty}\text{d}R\,R\,e^{-R^2}e^{-2 \frac{\Lambda^2}{g^2} (R^2-1)^2}}\ .
\end{split}
\end{equation}
The integral can be done by converting it to a Riemann sum and calculating each term in the sum numerically.

In Figure \ref{fig:vacuum_survival_gm2}, we plot the vacuum survival probability at $g^{-2}=10$ as a function of time, for various values of the radial cutoff $\Lambda$. We obtain very good agreement with exact results for time $t\lesssim 4$. Results for finite $\Lambda$ start to diverge from exact results for $t\gtrsim 4$ where larger values of $\Lambda$ are needed to maintain accuracy.

\begin{figure}[ht!]
    \centering
    \includegraphics[width=0.85\linewidth]{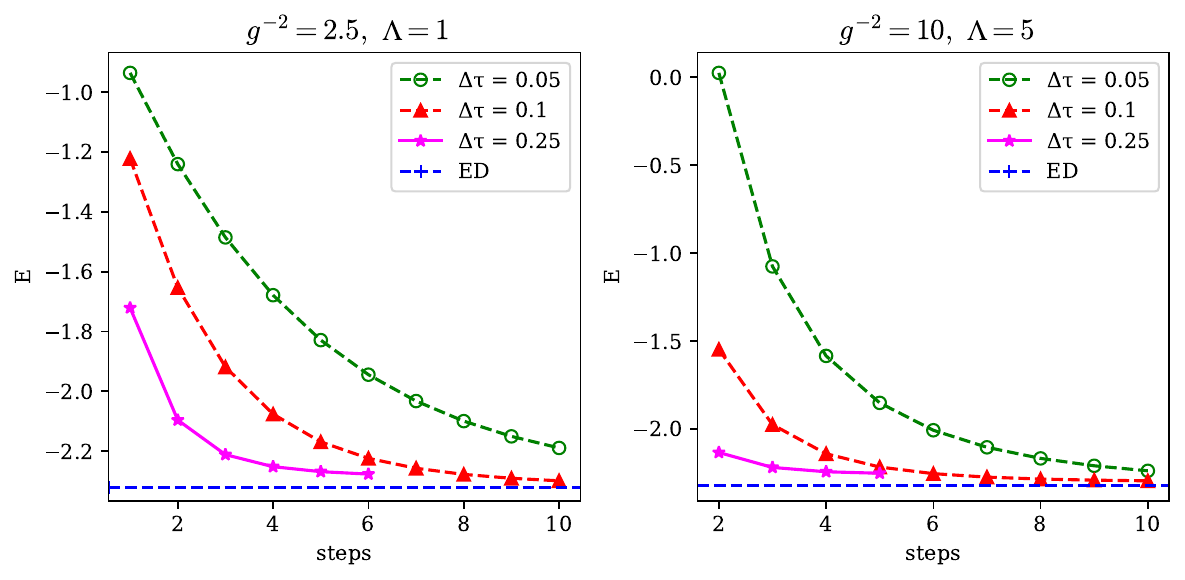}
    \caption{Energy of a single plaquette under QITE for different choices of the imaginary time increment $\Delta \tau$, coupling constant $g$ and radial cutoff $\Lambda$. The fermion mass is set at $m_0=1.5$ in both cases. The dashed blue line is the result obtained from Exact Diagonalization (ED).}
    \label{fig:tau_comp}
\end{figure}

\subsection{Ground state using QITE}\label{sec:gse_QITE}

Another method to compute the ground state energy is by using QITE, as discussed in Section \ref{sec:4}. To implement the procedure for a single plaquette, it is convenient to start with the initial
 state given in Eq.\ \eqref{eq:free_vacuum_uc}.
For each QITE step, we apply the non-unitary operator
\begin{equation}
    e^{- \Delta \tau H^{(Q)}} \approx U_\text{QITE} \equiv e^{-\Delta \tau H_E^{(Q)}} e^{-\Delta \tau H_B^{(Q)}} e^{-\Delta \tau H_M^{(Q)}} e^{-\Delta \tau H_K^{(Q)}}\,,
\end{equation}
see Eqs.\ \eqref{eq:HQ0} and \eqref{eq:HQ1}, using the quantum circuits discussed in Section \ref{sec:4}. After $S$ QITE steps, we arrive at the estimate of the ground state
\begin{equation}
  \ket{\psi_\text{ground}} = \frac{  (U_\text{QITE})^S \ket{V}}{\sqrt{\bra{V} (U_\text{QITE})^{2S} \ket{V}}}\,,
\end{equation}
and the corresponding estimate of the ground state energy
\begin{equation}
    \epsilon_0 = \bra{\psi_\text{ground}} H^{(Q)} \ket{\psi_\text{ground}} = \frac{\bra{V} H^{(Q)} (U_\text{QITE})^{2S} \ket{V}}{\bra{V} (U_\text{QITE})^{2S} \ket{V}}\,.
\end{equation}
Ideally, we would take $\Delta \tau \to 0$ and the number of steps $S \to \infty$. However, due to finite resources, a realistic implementation would require a finite $S$ and $\Delta \tau$ chosen carefully to balance the requirement that $\Delta \tau$ be sufficiently small (due to Trotterization) and the total evolution time $S\Delta \tau$ be large enough to allow for the decay of the system to a good approximation of the actual ground state.

\begin{figure}[t]
    \centering
    \includegraphics[width=0.7\linewidth]{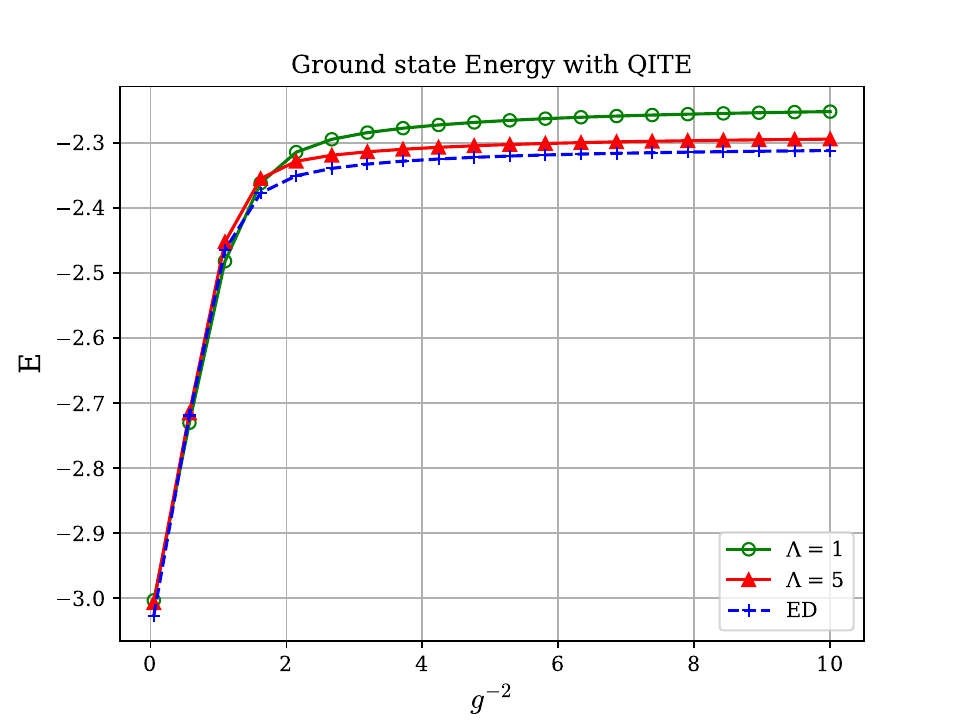}
    \caption{Estimate of the ground state energy of a single plaquette \emph{vs.}\ the coupling constant using QITE with $S=10$ steps and decay time increment $\Delta \tau = 0.1$. Fermion mass is set at $m_0=1.5$. Plots are shown for two different values of the radial cutoff $\Lambda = 1,5$ and ED.}
    \label{fig:QITE_vs_ExactD_plus_break_corrected}
\end{figure}

In Figure \ref{fig:tau_comp}, we plot the energy of a single plaquette undergoing imaginary time evolution for representative values of the coupling constant and radial cutoff $\Lambda$, and fermion mass $m_0 = 1.5$. As expected, smaller imaginary time increments $\Delta\tau$ require more QITE steps to converge. However, for $\Delta\tau = 0.25$, even after $S=4$, the energy is remarkably close to the ground state energy. The best results are obtained for $\Delta\tau = 0.1$ after $S=10$ QITE steps. They are shown in Figure \ref{fig:QITE_vs_ExactD_plus_break_corrected} for various values of the coupling constant. We obtain good agreement with exact results.

\begin{figure}[t]
    \centering
    \includegraphics[width=0.7\linewidth]{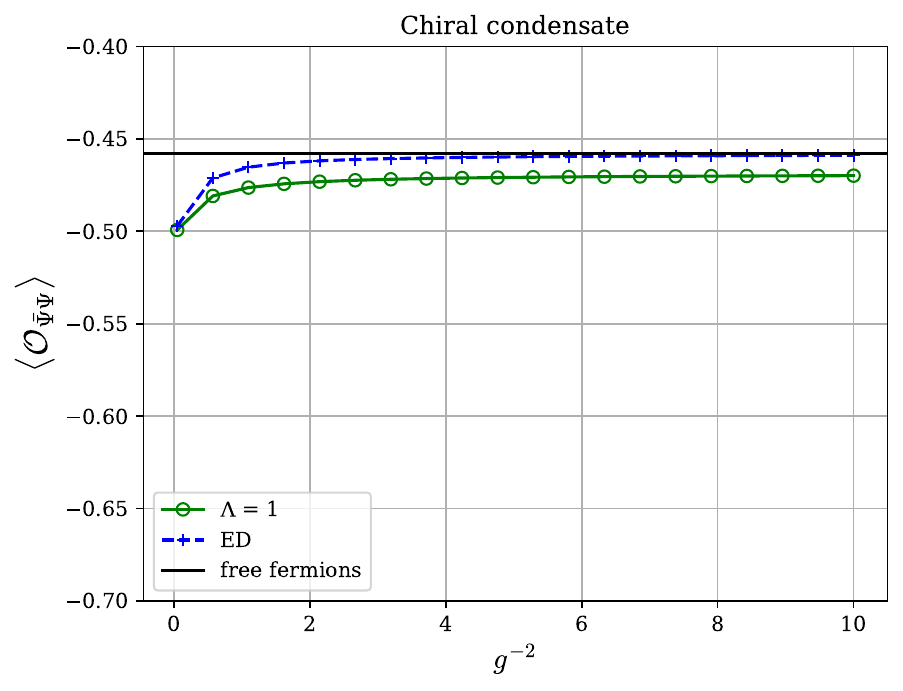}
    \caption{Vacuum expectation value of the chiral condensate for a single plaquette \emph{vs.}\ the coupling constant using QITE with $S=10$ steps, decay time increment $\Delta \tau =0.1$, radial cutoff $\Lambda = 1$, and fermion mass $m_0=1.5$, compared with ED and free fermions (Eq.\  \eqref{eq:chi_cnd}).}
    \label{fig:Chi_polar_mass_1.5}
\end{figure}

Next, we compute the vacuum expectation value of the chiral condensate,\footnote{It is worth noting that the chiral condensate is also relevant to condensed matter physics.
For example, in Ref.\ \cite{PhysRevB.79.165425, ARAKI20111408}, it is argued that the low-energy description of graphene resembles massless QED. Therefore, the chiral condensate of lattice QED in 2+1 dimensions was used as an order parameter to probe the semimetal-insulator transition in graphene, albeit at strong couplings and small masses.} 
\begin{equation}\label{eq:chi_cnd0}
   \mathcal{O}_{\bar{\Psi} \Psi} \equiv \frac{1}{4}\sum_{i=1}^{4}(-1)^{i+1}  \bar{\Psi}_i\Psi_i.
\end{equation}
Using QITE, we obtain the estimate for the vacuum expectation value,
\begin{equation}
    \braket{\mathcal{O}_{\bar{\Psi} \Psi}} = \bra{\psi_\text{ground}} \mathcal{O}_{\bar{\Psi} \Psi} \ket{\psi_\text{ground}} = \frac{\bra{V} (U_\text{QITE})^{S}\mathcal{O}_{\bar{\Psi} \Psi} (U_\text{QITE})^{S} \ket{V}}{\bra{V} (U_\text{QITE})^{2S} \ket{V}}
\end{equation}
Our results are shown in Figure \ref{fig:Chi_polar_mass_1.5}. We used QITE with $S=10$ steps, decay time increment $\Delta \tau =0.1$, radial cutoff $\Lambda = 1$, and fermion mass $m_0=1.5$. We obtained good agreement with ED. We In the weak coupling limit, we recover the free fermion chiral condensate, which is given by the analytic expression
\begin{equation}\label{eq:chi_cnd}
   \braket{\mathcal{O}_{\bar{\Psi} \Psi}}_{\rm free\, fermion\, ground\, state}  = -\frac{1}{4}\left(1+\frac{m_0}{\sqrt{1+m_0^2}}\right).
\end{equation}
In the strong coupling limit ($g^{-2}\ll 1$), the electric contribution dominates, and the fermions are effectively static charges. Therefore, we expect the chiral condensate to approximate the expression \eqref{eq:chi_cnd} in the limit $m_0\to\infty$, i.e., $ \braket{\mathcal{O}_{\bar{\Psi} \Psi}} \to -\frac{1}{2}$. This is in agreement with our results depicted in Figure \ref{fig:Chi_polar_mass_1.5}.

\section{Conclusion and outlook}
\label{sec:7}

We have formulated and analyzed a hybrid qubit–qumode framework for simulating U(1) lattice gauge theory coupled to fermionic matter in 2+1 dimensions, providing an explicit link between the field-theoretic Hamiltonian and experimentally accessible quantum-hardware operations. By encoding fermionic matter fields in discrete variables (qubits) and gauge fields in continuous bosonic modes (qumodes), the approach captures the full structure of the theory without introducing an artificial truncation of the gauge sector. Two compactness-enforcement strategies were developed: a squeezing-based projection that restricts qumode amplitudes to the compact manifold, and a Hamiltonian penalty term that confines the continuous variables dynamically. Both methods reproduce the correct gauge-invariant spectrum and are compatible with currently available hybrid platforms.

The electric, magnetic, and matter–gauge interaction terms of the lattice Hamiltonian were decomposed into a universal set of qubit–qumode gates composed of Gaussian, non-Gaussian, and conditional displacement operations. This establishes a concrete and hardware-realizable path for simulating Abelian gauge dynamics on hybrid devices such as trapped ions and superconducting circuits. Scaling estimates indicate polynomial growth of resources with lattice size, suggesting that few-plaquette systems lie within the reach of near-term technology. Using a continuous-variable extension of the QITE algorithm, we demonstrated that hybrid imaginary-time dynamics can efficiently prepare ground states and preserve gauge invariance during evolution.

The hybrid architecture presented here merges the strengths of discrete- and continuous-variable computations and opens a path toward scalable simulations of quantum field theories. Future work will extend the present formulation to non-Abelian gauge groups, explore hybrid variational and error-mitigation strategies, and test the methods on emerging hybrid quantum processors. The combination of theoretical rigor and experimental feasibility makes hybrid qubit–qumode simulation a promising paradigm for realizing quantum field dynamics in the laboratory. Future work will also extend this hybrid framework to non-Abelian gauge groups and explore real-time dynamics on fault-tolerant architectures.


\acknowledgments
We would like to thank Matt Grau for helpful discussions. We acknowledge support by DOE ASCR funding under the Quantum Computing Application Teams Program, NSF award DGE-2152168, and DOE Office of Nuclear Physics, Quantum Horizons Program, award DE-SC0023687. The research was supported the DOE, Office of Science, Office of Nuclear Physics, Early Career Program under contract No. DE-SC0024358 and DE-SC0025881. The authors would like to thank Stony Brook Research Computing and Cyberinfrastructure, and the Institute for Advanced Computational Science at Stony Brook University for access to the SeaWulf computing system, made possible by grants from the National Science Foundation ($\#$1531492 and Major Research Instrumentation award $\#$2215987), with matching funds from Empire State Development’s Division of Science, Technology and Innovation (NYSTAR) program (contract C210148).

\bibliographystyle{utphys.bst}
\bibliography{bibliography}

\end{document}